\documentclass[11pt,a4paper]{article}

\usepackage{jheplike}
\usepackage{amsmath,amssymb}
\usepackage{epsfig,graphicx}
\usepackage{subfig}
\usepackage{graphicx}
\usepackage{rotating}
\usepackage{cancel}
\usepackage{bm}
\usepackage{color}
\usepackage{comment}
\usepackage{psfrag}
\usepackage{enumerate}
\usepackage{overpic}

\renewcommand\[{\left[}

\newcommand{\exclude}[1]{}

\def\be{\begin{equation}}
\def\ee{\end{equation}}

\numberwithin{equation}{section}
\title{
\vspace{2.5cm} 
\Large{\textbf{Musings on cosmological relaxation and the hierarchy problem\vspace{0.5cm}}}}

\author[a]{Joerg Jaeckel,\note[a]
{\href{mailto:jjaeckel@thphys.uni-heidelberg.de}
{jjaeckel@thphys.uni-heidelberg.de}}}
\author[b]{Viraf M. Mehta,\note[b]
{\href{mailto:mehta@thphys.uni-heidelberg.de}
{mehta@thphys.uni-heidelberg.de}}}
\author[c]{and Lukas T.~Witkowski\note[c]
{\href{mailto:l.witkowski@thphys.uni-heidelberg.de}
{l.witkowski@thphys.uni-heidelberg.de}}}

\affiliation{Institute for Theoretical Physics, University of Heidelberg, \\ 
Philosophenweg 19, 69120 Heidelberg, Germany \vspace{0.1cm}}

\date{}

\abstract{
Recently Graham, Kaplan and Rajendran~\cite{Graham:2015cka} proposed cosmological relaxation as a mechanism for generating a hierarchically small Higgs vacuum expectation value. Inspired by this we collect some thoughts on steps towards a solution to the electroweak hierarchy problem and apply them to the original model of cosmological relaxation~\cite{Graham:2015cka}. To do so, we study the dynamics of the model and determine the relation between the fundamental input parameters and the electroweak vacuum expectation value. Depending on the input parameters the model exhibits three qualitatively different regimes, two of which allow for hierarchically small Higgs vacuum expectation values.
One leads to standard electroweak symmetry breaking whereas in the other regime electroweak symmetry is mainly broken by a Higgs source term. While the latter is not acceptable in a model based on the QCD axion, in non-QCD models this may lead to new and interesting signatures in Higgs observables.}

\begin{document}
\maketitle

\section{Introduction}
Run I of the LHC has seriously challenged the known approaches for solving the hierarchy problem. Most notably, supersymmetry and extra dimensions have not yet been found. The scale where they can solve the hierarchy problem has increased significantly and is challenging our notions of naturalness. While it is still too early to draw a definite conclusion (and hopes for discovery at Run II are high), it is nevertheless timely to think of new approaches to the hierarchy problem. Having a wide range of (new) solutions to this problem can serve to heighten and widen our perception of what to look for at the LHC and to also device complementary search strategies.

Indeed, recently an interesting new approach for making progress towards solving the hierarchy problem 
has been proposed~\cite{Graham:2015cka} (for previous work along these lines see~\cite{Abbott:1984qf,Dvali:2003br,Dvali:2004tma}). Further improved models based on this idea have also been constructed~\cite{Espinosa:2015eda, Hardy:2015laa, Patil:2015oxa, Antipin:2015jia}.
The hierarchy problem is widely accepted as a very difficult problem and, consequently, progress might only be possible in small steps. One strategy thus amounts to solving only certain aspects of the hierarchy problem, thereby perhaps providing us with a piece of the puzzle that is the whole problem. This is also the path taken in~\cite{Graham:2015cka} which attempts to solve the so-called ``technical hierarchy problem''. 

In this note we want to examine the model of~\cite{Graham:2015cka} and discuss its merits as a piece of the puzzle.
To do that we first set out a number of small steps that we think may help solving the hierarchy problem.
We then study the dynamics of the model of~\cite{Graham:2015cka} and determine its behavior in different regions of parameter space. Given our findings, we then discuss how the model of~\cite{Graham:2015cka} relates to the hierarchy problem.

\subsection*{Making progress towards solving the hierarchy problem}
What can be considered progress towards solving the hierarchy problem? Here we want to discuss this issue in the context of an effective description valid up to a scale, $\Lambda$, that we want to embed into a more complete and fundamental theory.

The hierarchy problem is closely related to the issue of fine-tuning and is, in essence, a question about when a number can be ``naturally'' small and where this small number originates from. One immediate observation is that we can only meaningfully talk about ``smallness'' for {\emph{dimensionless}} numbers.
In theories like the Standard Model (SM), there are many parameters from which we can form dimensionless numbers, e.g.~ratios of fermion masses. Some of them, such as the ratio of the electron to the top mass, are indeed worryingly small and one may desire an explanation.

In the context of the hierarchy problem the question applies to very specific dimensionless numbers. In particular, we do not want to have to use any unexplained small values for parameters at the fundamental level, i.e.~at the UV scale.
Absent a solution in a UV complete theory, we can only apply this requirement to our effective theory.  Therefore, we would like to apply it to all dimensionless coupling constants and ratios between dimensionful couplings and powers of the scale $\Lambda\gg v=246\,{\rm GeV}$. 

Let us begin by giving an overview of the possible paths via which we can make progress. The first category of progress is the straightforward improvement of the {\bf smallness/tuning} of the numbers themselves. A model can be considered an improvement over the SM if no or only little tuning of its parameters is necessary at the UV scale. In particular, one can ask:
\begin{enumerate}
\item{} Does the model require dimensionless quantities at the UV scale $\Lambda$ that are $\ll 1$? 
\item{} If yes, is the product of all small dimensionless parameters $< v^2/\Lambda^2$? 
\end{enumerate}
Progress is made if the answer to any one of the above questions is negative.

Since we know (or assume) that the theory is not complete, we can ask additional questions. In particular, we can look at what happens when, in one form or another, we try to extend the model. We could do this by just considering a higher value of the UV scale, or by the inclusion of new degrees of freedom. A model is then attractive, from the point of view of the hierarchy problem, if it exhibits a mild  {\bf scaling behavior} and has nice {\bf embedding features}.  So, we can ask:
\begin{enumerate}
\setcounter{enumi}{2}
\item{} If we change $\Lambda$ to a larger scale $\Lambda'$ can we keep $v$ (or any other relevant low energy parameter) fixed without having to
adapt any UV parameters by (large) powers of $\Lambda'/\Lambda$? 
\item{} If adaptation of the parameters is necessary, is it less than $(\Lambda'/\Lambda)^2$?
\item{} Is there a prescription for extending the model by additional fields with masses $M>\Lambda$ without requiring the original parameters to be changed by orders of magnitude?
\end{enumerate}
Further, as extremely small non-vanishing numbers are difficult to explain, one could aim to replace them by zeros. In general, one can try to 
improve the {\bf structure/parametrization of the Lagrangian to facilitate embedding}.
We therefore would also like to consider the following options:
\begin{enumerate}
\setcounter{enumi}{5}
\item{} Can (some of) the small numbers in the model be replaced by zeros, while keeping low energy parameters of interest ($v$ in the case of the hierarchy problem) intact? (Note, that we do not require an explanation for these zeros.)
\item{} Vice versa, can we explain a vanishing or extremely small ($\ll v$) Higgs mass parameter?
\item{} Can the desired smallness of the Higgs be achieved by choosing one (or at most only a few) of the dimensionless input parameters to be small? For example, is the Higgs vev proportional to a coupling parameter of the theory (whose smallness then gives the small Higgs vev)? 
\item{} Is the Lagrangian of the theory generic with respect to all (approximate) symmetries of the system, i.e.~are all terms that are allowed by symmetries present in the Lagrangian and are they of the same order of magnitude? In case of approximate symmetries, do all symmetry breaking terms have the same size?
\end{enumerate}

While the above list is not exhaustive, we think it can be used as guidance towards a solution to the hierarchy problem. In particular, the nine points introduced above can be used to assess to what extent a model can be considered an improvement over the SM with respect to the hierarchy problem. For details on this we refer readers to section \ref{sec:smallsteps}. Ultimately, we will apply these criteria to the paradigm of `Cosmological Relaxation' \cite{Graham:2015cka}. Readers mainly interested in the model of `Cosmological Relaxation' may skip directly to section \ref{sec:dynamics}.

\section{Small Steps towards a solution to the Hierarchy Problem}
\label{sec:smallsteps}
In this section we discuss possible pathways towards solving the hierarchy problem in more detail. To begin, we make our notion of tuning more precise. Using this understanding of tuning we then discuss the potential improvements in more detail. Where possible, we also try to give simple examples.

\subsection*{Fine-tuning}
To set the stage for a discussion of the steps listed in the introduction, let us first state what we consider as fine-tuning (for an interesting discussion and review of naturalness and fine-tuning see~\cite{Giudice:2008bi}). 

\begin{figure}[t]
\includegraphics[width=1.0\textwidth]{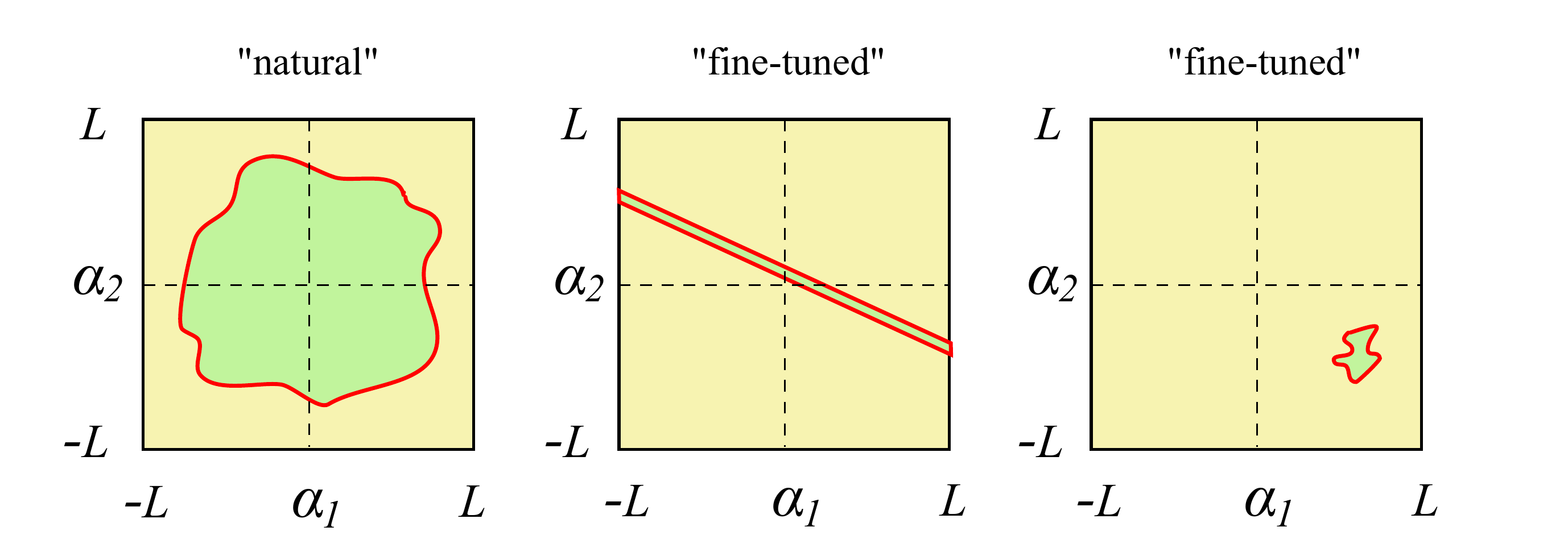}
\caption{Sketch of what appears to be natural versus situations one may consider fine-tuned.}
\label{finetuningsketch}
\end{figure}

Consider the space of the fundamental input parameters, i.e.~all parameters that are consistent with the symmetries and field content of the system. A measure of fine-tuning is then given by the fraction of parameter space reproducing qualitatively similar physics to what is observed (see also~\cite{Ciafaloni:1996zh}). If this fraction is very small the model is fine-tuned\footnote{This is essentially the notion of wanting an explanation for any small numbers~\cite{Dirac:1937ti,Dirac:1938mt}.} (see Fig.~\ref{finetuningsketch}).

Let us look at a parameter that is, in some sense, small (or otherwise peculiar) at the low energy scale,\footnote{It is, for example, trivial to generalize to the case where a low energy parameter takes some very large value. But more generally one could simply require that the low energy parameter should sustain physics that is qualitatively similar to what we observe.} e.g.~the Higgs vev, which we will use, henceforth, for concreteness, but our arguments hold more generally, e.g.~for the case of the flavour problem.
We can now ask ourselves the following: Which fraction of the fundamental parameter space (at some high energy scale) leads to a Higgs vev of a size that is observed or smaller?\footnote{With the latter choice, we avoid any tuning that would arise from simply having measured the vev to a high degree of precision. Nevertheless, in some cases one may also want to impose a restriction on the lower values, e.g.~when the qualitative behavior changes. For example, for a Higgs vev smaller than the QCD scale, electroweak symmetry would be broken by the condensates that are usually responsible for chiral symmetry breaking. A mechanism that produces such small Higgs vevs for nearly all of the available parameter space may also not be satisfactory. An alternative, and perhaps more precise possibility, would be to take the volume of parameter space for which the measurable observables at low energies lie within a reasonable factor, say a factor of $2$, of the observed values.} We would then measure the fine tuning in the following way. Let $\alpha_{1},\ldots,\alpha_{n} \in \{ -L ,L \}$ with $L \sim \mathcal{O}(1)$ be the space of all fundamental, {\bf scaled to be dimensionless}, input parameters consistent with the symmetries and field content of our system. The total volume in parameter space is then,
\begin{equation}
{\rm Vol}_{\rm total}=\prod_{i=1}^{n} \int^{L}_{-{L}} d\alpha_{i}.
\end{equation}
Only a volume,
\begin{equation}
{\rm Vol}_{\rm v\leq v_{0}}=\prod_{i=1}^{n} \int^{L}_{-{L}} d\alpha_{i}\,\theta(v_{0}-v(\alpha_{i}))
\end{equation}
would lead to a vev that is smaller than the physical value $v_{0}$. The ratio then corresponds to a measure of the amount of fine-tuning required:
\begin{equation}
\label{eq:Ftune}
F=\frac{{\rm Vol}_{\rm total}}{{\rm Vol}_{\rm v\leq v_{0}}} \ .
\end{equation}

A high degree of fine-tuning can manifest itself in different ways. For example, the model is highly fine-tuned if one parameter has to take very specific values. Alternatively, a high level of fine-tuning is also observed if two or more parameters are required to take somewhat specific values. See also Fig.~\ref{finetuningsketch}.

Comparing theories with different numbers of free parameters is always rather difficult. However, the amount of fine-tuning as defined above will typically increase with the number of parameters. This can be easily understood: Unless additional parameters are free to take values in their entire range, they will cause the ratio \eqref{eq:Ftune} to decrease, thereby increasing the fine-tuning.

This could be dealt with in two ways:
\begin{enumerate}[i)]
  \item This increase in the fine-tuning measure $F$ with the number of paramaters could actually be taken as a desirable feature. It penalises complicated or `baroque' constructions with many parameters. One could therefore take $F$ as some form of combined measure of fine-tuning and `baroqueness', which one would like to minimize (see~\cite{resonaances}).  
  \item One could also simply multiply by a factor $\sim 1/L^n$
to account for this (which, of course, entails a choice of $L$).
\end{enumerate}

\subsection*{Progress towards solving the hierarchy problem}

Equipped with the picture of fine-tuning above, let us now proceed to a discussion of our previous collection of steps that may help with solving the hierarchy problem. \\[0.1cm]
 
\noindent{\bf Ad 1) and 2):} It is clear that any viable model that has no small numbers does indeed solve the hierarchy problem up to a scale,
$\Lambda$. However, even if the model requires small parameters, improvement compared to the SM can be achieved if these parameters are less tuned than the equivalent tuning of $v^2/ \Lambda^2$ in the SM. Thus, we would consider it a step forward if the small parameters do not have to be as small as $v^2/\Lambda^2$. 

However, one has to be careful once there is more than one small parameter: if the product of all small parameters is again of the order of $v^2/\Lambda^2$ or smaller, it remains questionable whether any progress has been achieved. The level of tuning required, according to our definition, is then as high or even higher than in the SM (cf.~the left hand side of Fig.~\ref{finetuningsketch}).\\[0.1cm]

\begin{figure}[t]
\begin{center}
\includegraphics[width=0.66\textwidth]{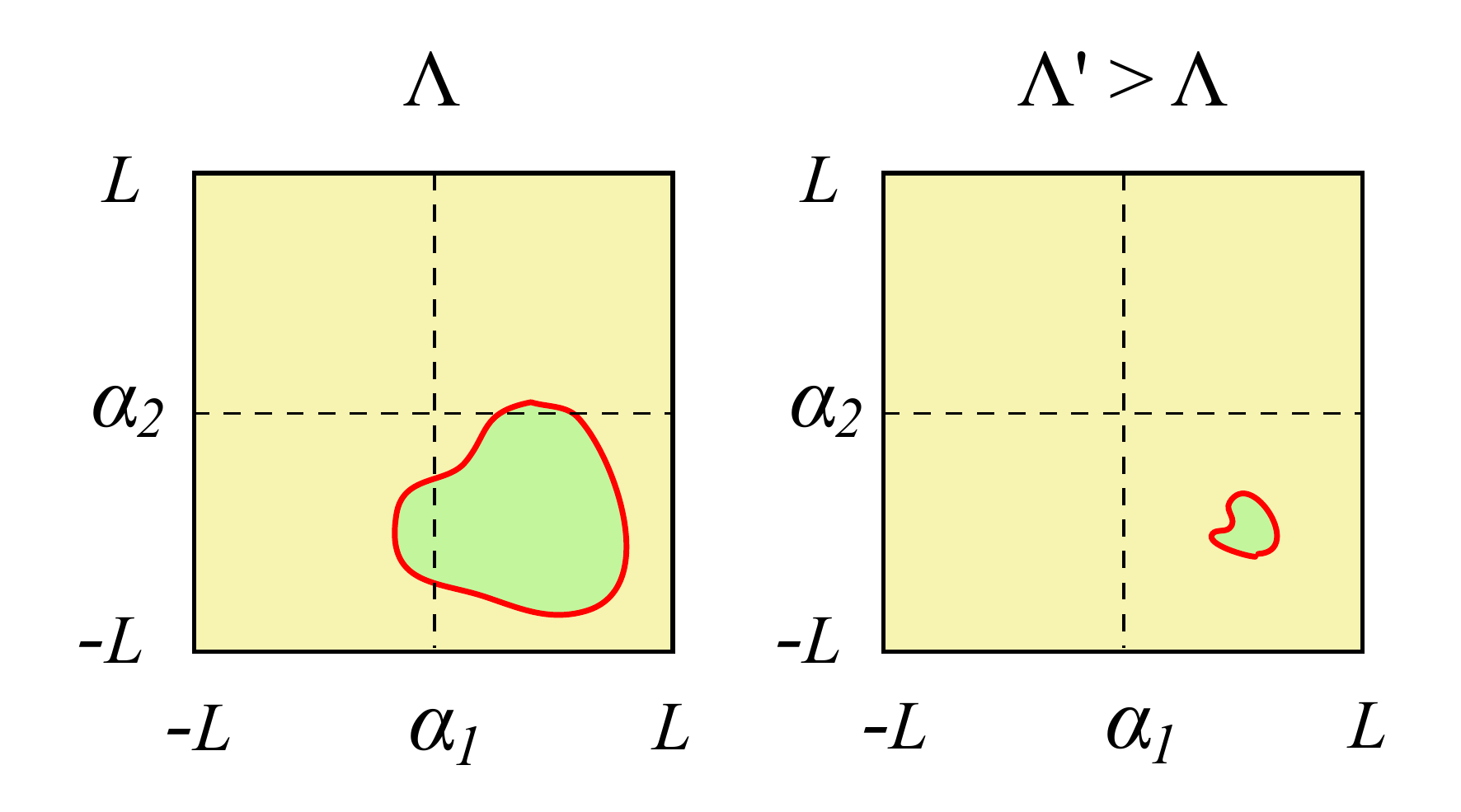}
\end{center}
\caption{When trying to increase the cutoff scale, fine-tuning may become worse.}
\label{scaling}
\end{figure}

\noindent{\bf Ad 3) and 4):} 
When trying to embed our theory, an obvious question arises: what happens to the allowed (dimensionless) parameter space when we increase the cutoff scale as $\Lambda \rightarrow \Lambda' > \Lambda$ (see Fig.~\ref{scaling})? If the allowed parameter space remains constant the amount of fine-tuning is independent of the cutoff scale. On the other hand, if the allowed parameter space shrinks, fine-tuning worsens as we try to approach higher energy scales. This is shown in Fig.~\ref{scaling} and corresponds to the situation of the Higgs in the SM (as long as the situation remains perturbative; if the anomalous dimension, for some reason, approaches $2$ it could remain constant~\cite{Wetterich:1983bi}).
 
To give an example of this, let us consider the SM extended by a right handed neutrino. In particular, let us require the mass of the right handed Majorana neutrino measured at the low energy scale $k$ to be equal to that of an electron: $M_{R}(k\approx 0)=m_{e}(k\approx 0)$. 
As the fundamental dimensionless input parameters we have the Yukawa coupling $Y(\Lambda)$ and the Majorana mass $\epsilon_{R}(\Lambda)=M_{R}(\Lambda)/\Lambda$. 

Here we want to study how the input parameters have to be adapted for different UV scales, $\Lambda$. As an example, let us take $\Lambda=v$ and $\Lambda'=100\Lambda=100v$.
To get the correct electron mass in a theory with cutoff $\Lambda$ requires $Y(\Lambda=v)\sim 10^{-6}$ and $\epsilon_{R}(\Lambda=v)\sim 10^{-6}$. 
However, if we increase the UV scale to $\Lambda'$ (while keeping the vev of the Higgs field $v$ constant),\footnote{This, of course, also requires tuning, but our example demonstrates that a small Majorana mass would be an independent tuning which is more severe than say an additional chiral fermion.} the Yukawa coupling only has to be adapted logarithmically to preserve the observed low scale mass and we still have $Y(\Lambda')\sim 10^{-6}$.
For the Majorana mass to remain the same we now need $\epsilon_{R}(\Lambda')\sim \epsilon_{R}(\Lambda)\Lambda/\Lambda'\sim 10^{-8}$.
So for the Majorana case significant adaptation is required.
In a situation akin to the electron case, it seems that it is easier to establish a large hierarchy\footnote{As an elaboration of our example, one could make
the argument that for a theory valid up to the Planck scale $\Lambda\sim 10^{18}$~GeV a TeV scale Majorana neutrino is actually somewhat more tuned than 
a $0.1$~eV Dirac neutrino with imposed lepton number conservation, since $M_{R}/\Lambda\sim 10^{3}\,{\rm GeV}/(10^{18}\,{\rm GeV})\sim 10^{-15}$ 
whereas $Y\sim m_{\nu}/v\sim 10^{-12}$.
}.

The difference between the two cases is of course that one of the input parameters has a non-vanishing mass dimension.
In the UV, their RG behaviour with respect to the scale $k$ is of the form,
\begin{equation}
\partial_{t}x(t)=-d_{x}x(t),\qquad t=\log(k).
\end{equation}
We then have for the dimensionless parameters $Y(k)$ and $\epsilon_{R}(k)=M_{R}(k)/k$
\begin{equation}
d_{Y}\approx 0,\qquad d_{\epsilon_{R}}\approx 1 \ ,
\end{equation}
i.e.~the scaling dimensions in the UV differ. In the above example, and in many perturbative situations, this simply means that we want to have truly dimensionless input parameters at the UV scale\footnote{If we go beyond the perturbative regime and consider a potential non-trivial UV fixed point of the renormalization group it simply means that the parameters with UV scaling dimension 0 seem preferable from the hierarchy point of view.}. One such example, in the context of the electroweak hierarchy problem, is technicolor~\cite{Weinberg:1975gm,Susskind:1978ms,Dimopoulos:1979es,Eichten:1979ah}. Since there are no fundamental scalars or other dimensionful quantities (in the UV), 
all couplings have a scaling dimension of zero. 

However, an improvement over the SM is already achieved when the scaling dimension (e.g.~via a significant anomalous dimension in the UV) of the UV input
parameters is reduced. For the Higgs field mass parameter in the SM, the dimensionless UV input quantity is
\begin{equation}
\label{pick}
\epsilon_{H}(\Lambda)\sim \frac{m^{2}_{H}}{\Lambda^2} \ .
\end{equation}
The dimensional running (i.e.~neglecting small anomalous dimensions) of $\epsilon_{H}=m^{2}_{H}/k^2$ 
is thus given by
\begin{equation}
\partial_{t}\epsilon_{H}=-d_{\epsilon_{H}}\epsilon_{H},\quad d_{\epsilon_{H}}=2.
\end{equation}
So in that sense, progress compared to the SM is already made when the necessary input parameters have scaling dimension
less than $2$.

In addition, both couplings discussed in the above example are equally natural in the context of 't Hooft's definition~\cite{tHooft:1979bh}.
In the first case, a vanishing Yukawa coupling allows one to do independent right handed chiral rotations for the electron, while in the case of a vanishing Majorana neutrino mass the corresponding lepton number is conserved. Thus, while it is technically natural to have both quantities small, a small electron mass is nevertheless preferable from a tuning point of view. 

So far we have only considered situations where the measured quantity of interest in the IR is essentially the same as the input parameter in the UV. Yet, in many cases it depends on several of the input parameters of the theory. Let us consider the simple example of a scalar field with a potential
\begin{equation}
\label{scalarfieldex}
V(\phi)=\frac{1}{2}m^2\phi^2-\kappa\phi.
\end{equation}
In this case, the vev is given by
\begin{equation}
\langle \phi\rangle =\frac{\kappa}{m^2}.
\end{equation}
We want to consider a situation where we have a small fixed $\langle \phi \rangle$.

Now we have to look at the scaling of both parameters. In absence of any knowledge regarding the origin of the parameters $\kappa$ and $m^2$, we have to assume that they scale according to their na\"{i}ve dimension. The dimensionless input parameters are,
\begin{equation}
\alpha_{1}=\frac{m^2}{\Lambda^2},\qquad \alpha_{2}=\frac{\kappa}{\Lambda^3}.
\end{equation}
For the vev we therefore find\footnote{Since there are no interactions, the parameters $m^2$ and $\kappa$ in the infrared are the same as in the UV.},
\begin{equation}
\langle\phi\rangle=\frac{\alpha_{2}}{\alpha_{1}}\Lambda.
\end{equation}
Increasing $\Lambda\to\Lambda'$, while keeping $\langle \phi\rangle=v$ fixed, we need to reduce $\alpha_{2}/\alpha_{1}$ by a factor
$\Lambda/\Lambda'$ and the fine-tuning increases accordingly.

However, the situation can change when we have further information. For example, imagine we have a mechanism that determines $\kappa$ from some independent scale $M$ such that $\kappa=M^3$. Most important, $M$ is fixed when $\Lambda$ is increased. In this case it is fairly straightforward to explain a very small $\langle \phi\rangle$. One just needs a large enough $\Lambda$. Alternatively, if there is a mechanism that produces a small and cutoff-independent $m^2$ (or if we have measured $m^2$ to be small) the situation may actually become worse.
It is thus important to know which parameters can be treated as free parameters.
\\[0.1cm]

\noindent
{\bf Ad 5:} So far we have simply and very na\"{i}vely increased the UV scale of the theory. A perhaps more meaningful
change to the theory occurs when we change the field content by including additional particles, possibly with a mass $M>\Lambda$. This is indeed what is happening when one ``integrates in'' degrees of freedom to embed the model in a theory with a higher UV cutoff (or in a UV complete theory).

Adding such additional fields is of interest as, by a suitable choice, one may improve the fundamental UV behavior of the Higgs. For example, in string theories, higher string modes actually render the theory finite. Here, however, we are not looking at such a complete theory but are trying to make progress in the context of an effective theory. Thus, we have to ask, what would make such an embedding easier and safer from a fine-tuning point of view?

One possibility is that there is a well defined prescription to add combinations of fields with mass $M>\Lambda$, coupled to the Higgs and other relevant fields (otherwise it is trivial), such that no (or little) adaptation of the 
UV input parameters of the model, without the fields, is needed.

The presence of a symmetry, as required by 't Hooft, usually gives exactly such a prescription. A famous example is, of course, supersymmetry~\cite{Fayet:1976et,Fayet:1977yc,Farrar:1978xj,Fayet:1979sa}~\cite{Ramond:1971gb,Neveu:1971rx,Gervais:1971ji,Golfand:1971iw,Volkov:1973ix,Wess:1974tw}. If we add a supersymmetric multiplet that does not introduce any additional sources of SUSY breaking (including spontaneous breaking), then the soft parameters of the Higgs and its vev are only changed by ${\mathcal{O}}(1)$ factors.
This holds, even if the multiplet in question is very heavy.

However, one has to be careful when breaking the symmetry.
In the case of SUSY, only so-called soft-breaking is allowed. This is such that no new sensitivities to the cut-off or the mass scales of new
particles are introduced. In particular,  to prevent quadratic divergences from reappearing, SUSY should only be broken by soft mass terms and soft couplings of three scalars~\cite{Dimopoulos:1981au,Witten:1981nf,Dine:1981za,Dimopoulos:1981zb,Sakai:1981gr,Kaul:1981hi} (see also~\cite{Martin:1997ns}). Note that couplings are more dangerous and, in general, only a subset of interactions is allowed, if any at all. 

This difference can also be seen in the example Eq.~\eqref{scalarfieldex} which features a $Z_{2}$ symmetry broken by the source term $\sim\kappa\phi$. Adding a term $\sim \xi \phi^3$ we get a correction to $\kappa$, $\Delta \kappa\sim \xi\Lambda^2$ that is proportional to the cutoff squared, indicating that the coupling $\xi$ does not correspond to a soft breaking. The same holds if we introduce couplings like $\phi\chi^2$ to a new field $\chi$.

We hence arrive at another principle which helps in defining progress towards a solution of the hierarchy problem: If we consider a theory with a small parameter that is technically natural (i.e.~setting this parameter to zero restores a symmetry), it is preferable for this symmetry to be broken softly. Otherwise we risk reintroducing problems when embedding the model in a more complete theory.
\\[0.1cm]

\noindent
{\bf Ad 6 and 7:} Perhaps it is simpler to explain an exact zero rather than a very small number. 
For example, in the Standard Model, we could ask whether we can set the Higgs mass parameter in the UV to be exactly zero
and still obtain a phenomenologically acceptable Higgs vev. This is the idea behind models based on the Coleman-Weinberg mechanism~\cite{Coleman:1973jx}.
Of course, in the SM this does not work, but can be made viable by adding an additional hidden sector (see, e.g.~\cite{Hempfling:1996ht,Meissner:2006zh,Chang:2007ki,Foot:2007as,Foot:2007iy,Meissner:2007xv,Iso:2009ss,Holthausen:2009uc,AlexanderNunneley:2010nw}).

While one can argue that there is no exact symmetry to justify this (scale invariance is anomalous; but one could still argue for classical scale invariance), it would still correspond to
a well defined prescription for the effective model~\cite{Bardeen:1995kv,Meissner:2007xv,Englert:2013gz}. This is clear in dimensional regularization because the RG equation for the Higgs mass parameter has a fixed point where the mass parameter vanishes (this holds similarly for other scalar field mass parameters). Even in explicitly scale-invariance-breaking regularization schemes, it corresponds to a selection of a well defined hypersurface in the space of all couplings (see also~\cite{Englert:2013gz}). It is at least possible that a selection of such a hypersurface may be easier to justify in a more complete theory than a very small 
parameter. 

Vice versa to 6), a mechanism that allows one to set problematic parameters to zero, or close to zero, may help. Indeed it could, for example, be the starting point for a model based on scale invariance.\\[0.1cm]

\noindent{\bf Ad 8 and 9:} Fine-tuning is fundamentally a problem of parametrization. Since we do not (yet) know the measure on the parameter space, 
one could always choose a non-linear parametrization that blows up the desired region in parameter space and/or shrinks the undesirable ones.
This can be illustrated by the example of QCD. Parametrizing the nucleon masses in terms of the QCD scale $\Lambda_{QCD}$, i.e.~we have $m_{nucl}\sim \Lambda_{\rm QCD}$, QCD looks horribly tuned since $\Lambda_{\rm QCD}/M_{P}\sim 10^{-19}\ll 1$ is required. However, in terms of the QCD coupling, we have,
\begin{equation}
\Lambda_{\rm QCD}\sim M_{P}\exp\left(-\frac{8\pi^2}{b_{0}g^2(M_{P})}\right),
\end{equation}
and $g\sim 1$ gives us the right QCD scale ($b_{0}$ is the $\beta$-function coefficient). No significant tuning is required anymore. Similarly, technicolor achieves such a rescaling for the electroweak vev\footnote{Indeed this rescaling has its root cause in the scaling dimension of $g$ being $=0$ in the UV. However, a vanishing scaling dimension is not always sufficient to address all issues of unnaturally small numbers. For example, the electron Yukawa coupling does have scaling dimension zero in the UV but we would still like to understand its value $\sim 10^{-5}$.}.

Importantly, such a rescaling seems easier to achieve if the quantity for which we have measured a small value is related in a simple manner 
to the fundamental input parameters, most notably by being proportional to some power of this input parameter. For example, in the case of QCD the nucleon mass is proportional to the QCD scale itself.

\begin{figure}[t]
\includegraphics[width=1.0\textwidth]{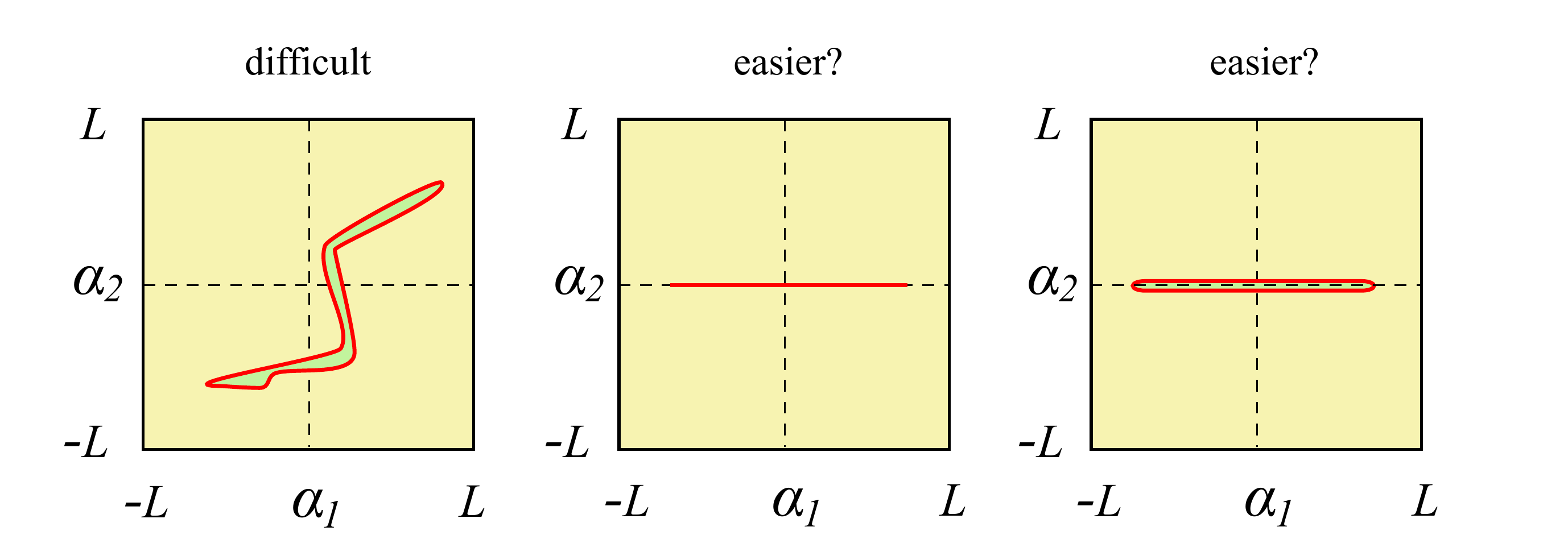}
\caption{Sketch of different ways in which a parameter region can be small. The middle and right panels show situations which are still ``tuned'' but one can hope for an easier embedding.}
\label{embeddingcomp}
\end{figure}

In contrast, such an embedding seems much more difficult if the relevant region in parameter space has a complicated, non-trivial shape -- potentially even depending on several input parameters in a non-trivial way. 
Indeed, this is often considered to be what makes the hierarchy problem especially hard: The Higgs vev (or the required negative Higgs field mass parameter) requires a very non-trivial cancellation between a combination of the different masses and coupling parameters.
This is schematically illustrated in Fig.~\ref{embeddingcomp}. In contrast, one could imagine that
it is easier to find an embedding that corresponds to the situation indicated in the middle or the right panel of Fig.~\ref{embeddingcomp}. In the middle panel, one parameter can be chosen to be exactly zero as suggested by step 6. Alternatively, step 8 suggests the parameter could be small but non-vanishing. Most importantly, it is one parameter that is small and not a complicated combination of parameters. In both cases, the allowed parameter region is small (or even vanishing) but it may be easier to find an embedding.

If the observed small parameter can be related to an approximate symmetry, as in 't Hooft's definition, such a simple relation can typically be achieved.
For example, the electron Yukawa breaks the chiral symmetry for the electron and the mass is, therefore, proportional to it.
It does not fully explain the smallness of the electron mass, but one can now look for a reparametrization of this single small parameter (e.g.~by making it exponential as in models with extra dimensions and brane-type models~\cite{Randall:1999ee,Randall:1999vf,ArkaniHamed:1999dc}; or models where the Yukawa coupling is generated via some (stringy-) instanton effect~\cite{Abel:2006yk, Cvetic:2009yh}; or by generating it from the power of some smallish parameter, as in a Froggatt-Nielsen mechanism~\cite{Froggatt:1978nt}).

Having a simple functional dependence of the desired small quantity on the fundamental input parameters at the UV scale -- preferably on one (still small) quantity -- is what we would associate with, what is often termed, ``a solution to the technical hierarchy problem''. We stress that this is taken to refer to a simple parametrization at the UV scale in some physically preferred basis and regularization at the high scale. Indeed, if the embedding is simple in all regularization schemes, it may be even easier to achieve an embedding.

Let us stress the importance of having a fixed parametrization and a fixed regularization.
Leaving the choice of parametrization and regularization free, it is indeed always possible to obtain a parametrization such that the physical Higgs vev in the SM is essentially proportional to some small renormalized input parameter at the UV scale~\cite{Wetterich:1983bi}. The reason for this is that
a vanishing Higgs field mass parameter corresponds to a second order phase transition and one can simply define a suitable input parameter 
that measures the distance to this phase transition (for details on how this works see Appendix \ref{sec:app}).
This same parameter also measures the explicit breaking of scale invariance in the SM.\footnote{While scale-invariance is anomalous, we think that anomalous breaking terms can, in many cases, be separated from non-anomalous ones. For example, it is widely accepted that the axion obtains its potential from a purely anomalous breaking. Gravity effects aside, no (large) explicit breaking is required just because the $U(1)$ symmetry is anomalous.
In any case, anomalous scale-invariance breaking effects in the SM from QCD, etc. are small.}
In this sense, one could argue that (approximately) 't Hooft's naturalness criterion is fulfilled with respect to scale invariance.
The question is then, however, why the parameter is small in a given physical regularization scheme and parametrization, i.e.~in a desired embedding, and in particular one also still has to explain the smallness of the number itself. One should keep in mind that the latter always has to be the ultimate goal.

To summarize, progress towards solving the hierarchy problem can be made by constructing a model where the Higgs vev has a simple functional dependence on one or few small parameters in a physically preferred basis.
When realizing a simplification of the embedding along these lines, one should, of course, try to avoid introducing additional problematic
choices and tunings. Specifically, one should ensure that the potential is generic in the sense that all parameters consistent with the symmetry should have the same order of magnitude. In case of an approximate symmetry, one should similarly require that all symmetry breaking terms are of the same order.

Finally, let us remark that while a generic potential makes an embedding definitely easier, it is not absolutely necessary. Non-perturbative effects and, in particular, anomalies, are known to produce non-generic potentials~\cite{Seiberg:1993vc}.

\subsubsection*{Summary}
In this section we collected a set of features of effective field theories that can be considered progress towards solving the hierarchy problem. Ultimately, we want to apply these criteria to the paradigm proposed by \cite{Graham:2015cka}. However, before we do so, we need to understand the model of \cite{Graham:2015cka} in some detail. Thus, by studying the dynamics of the model we will derive when and how a small Higgs vev can be generated.

\section{The Dynamics of Cosmological Relaxation}\label{dynamics}
\label{sec:dynamics}
The nature of the hierarchy problem lies in the relation between the fundamental parameters of a theory and the
observable quantities; of particular significance for this discussion is the electroweak scale. Here we investigate such relations for the model proposed in~\cite{Graham:2015cka}.

\subsection{Model setup}
To properly define our conventions, let us write down the potential for the Higgs and axion fields,
\begin{equation}
\label{eq:V}
V(\phi,h)=-\frac{1}{2}(M^2+g\phi)h^2-c_{1}gM^2\phi+\frac{c_{2}}{2}g^2\phi^2+\frac{\lambda}{4}h^4-\kappa |h|\cos\!\left(\frac{\phi}{f}\right)
\end{equation}
where $h$ is the Higgs field and is taken to be real, i.e.~it represents the vev, and $\phi$ is the axion field.

Note that, in comparison with~\cite{Graham:2015cka}, we have changed a few sign conventions in order to have the field $\phi$ rolling
from large negative values towards positive values.

$M$ denotes the UV scale up to which the model should be valid and $f$ is the axion decay constant. Moreover, we have explicitly written down the Higgs field dependence of the pre-factor of the axion potential. Following the conventions of \cite{Graham:2015cka} we set
\begin{equation}
\Lambda^4=\kappa |h|,
\end{equation}
with $\kappa$ being a model-dependent coupling. (In the model where $\phi$ is the QCD axion $\kappa$ is fixed to be $m^{2}_{\pi}f^{2}_{\pi}/v$.) Importantly, we explicitly included two constants, $c_{1}$ and $c_{2}$, to make the pre-factors in front of generic structures evident.
Finally, we also include the stabilizing quartic coupling for the Higgs field, as assumed in \cite{Graham:2015cka}.

The important idea of this model can be summarized by the following: Initially, one assumes that the axion $\phi$ starts at a negative field value, $\phi_i= - \mathcal{O}(2) M^2/g$. The Higgs field can take any value, but will quickly settle in its initial minimum at $h \approx 0$. Thus, for simplicity, we can take $h_i=0$. When Hubble friction becomes subcritical, the axion will start to evolve and roll towards larger field values. Once it crosses $\phi=\phi_c =-M^2/g$ the mass term for the Higgs becomes tachyonic and the Higgs potential takes on its usual Mexican-hat form. Thus, for $\phi > \phi_c$ the field point $h=0$ becomes unstable and the Higgs will fall into its new minimum at the bottom of the potential. 
This, in turn, has consequences for the axion evolution: Once $h \neq 0$ the term $\sim |h| \cos \phi/f$ will act as a set of periodic barriers for the axion. If these barriers are high enough, the axion evolution will stop. The Higgs will remain in its minimum at the bottom of the Mexican-hat with a vev, $\langle h \rangle$, which can, in principle, be much smaller than the cutoff of the theory $M$. This would solve the hierarchy problem in that $\frac{\langle h \rangle}{M}$ is sufficiently small.

In \cite{Graham:2015cka} a set of constraints on the model parameters was also specified. The constant $g$, the mass parameter for the field $\phi$, should satisfy
\be
\label{eq:slowroll1}
g < H \ ,
\ee
\be
\label{eq:slowroll2}
g < \frac{H^2}{M^2} M_{pl}\ ,
\ee
for $\phi$ to be rolling slowly.  One also requires the energy density in the axion sector to be subdominant to the energy density of the inflation sector, \textit{i.e.}  
\be
H>\frac{M^2}{M_{pl}}
\ee
Finally, for the evolution of $\phi$ to be dominated by classical rolling, rather than quantum `jumps', one has
\be
\label{eq:nojump}
H < \frac{V'}{H^2} \ \ \Rightarrow \ \ H < (g M^{2})^{1/3} \ .
\ee
We will ensure that the conditions \eqref{eq:slowroll1} - \eqref{eq:nojump} are satisfied in all the numerical examples studied in the remainder of our discussion. Since we are only interested in the mechanisms itself and not in the particular numerical values we will, however, not try to reproduce a realistic Higgs vev.

It will also be useful to introduce the shifted axion field
\begin{equation}
\label{eq:tildephi}
\tilde{\phi}=\phi- \phi_{c} \quad{\rm and}\quad \phi_{c}=-\frac{M^2}{g}.
\end{equation}
Thus, in terms of $\tilde{\phi}$ the potential is given by
\be
\label{eq:tildeV}
V(\tilde{\phi},h)=-\frac{1}{2}g \tilde{\phi} h^2-(c_{1}+c_2) gM^2 \tilde{\phi}+\frac{c_{2}}{2}g^2\tilde{\phi}^2+\frac{\lambda}{4}h^4-\kappa |h|\cos\!\left(\frac{\tilde{\phi} + \phi_c}{f}\right) +(c_1+\frac{c_2}{2})M^4 \ .
\ee 

\subsection{Single-field approximation}
\label{sec:single}
The typical mass scale for the axion, or axion-like particle, is going to be very small. Therefore, the dynamics for the Higgs field will evolve much faster than those of the axion.
We can assume, as a rough approximation, that the Higgs field takes its minimum position instantaneously and thus, may consider the effective potential, $V(\phi)$.

Moreover, we can assume that the cosine-potential only constitutes a small effect. We can then easily find the minimum for the Higgs,
\begin{equation}
\langle h\rangle=\bigg\{\begin{array}{cl}
\frac{\sqrt{g(\phi-\phi_{c})}}{\sqrt{\lambda}}=\frac{\sqrt{g\tilde{\phi}}}{\sqrt{\lambda}} & \phi\geq \phi_{c},\,\,\tilde{\phi}>0
\\
0 & \phi<\phi_{c},\,\,\tilde{\phi}<0
\end{array}\ .
\end{equation}

\begin{figure}[t]
 \subfloat[][]{
\begin{overpic}[width=0.46\textwidth]{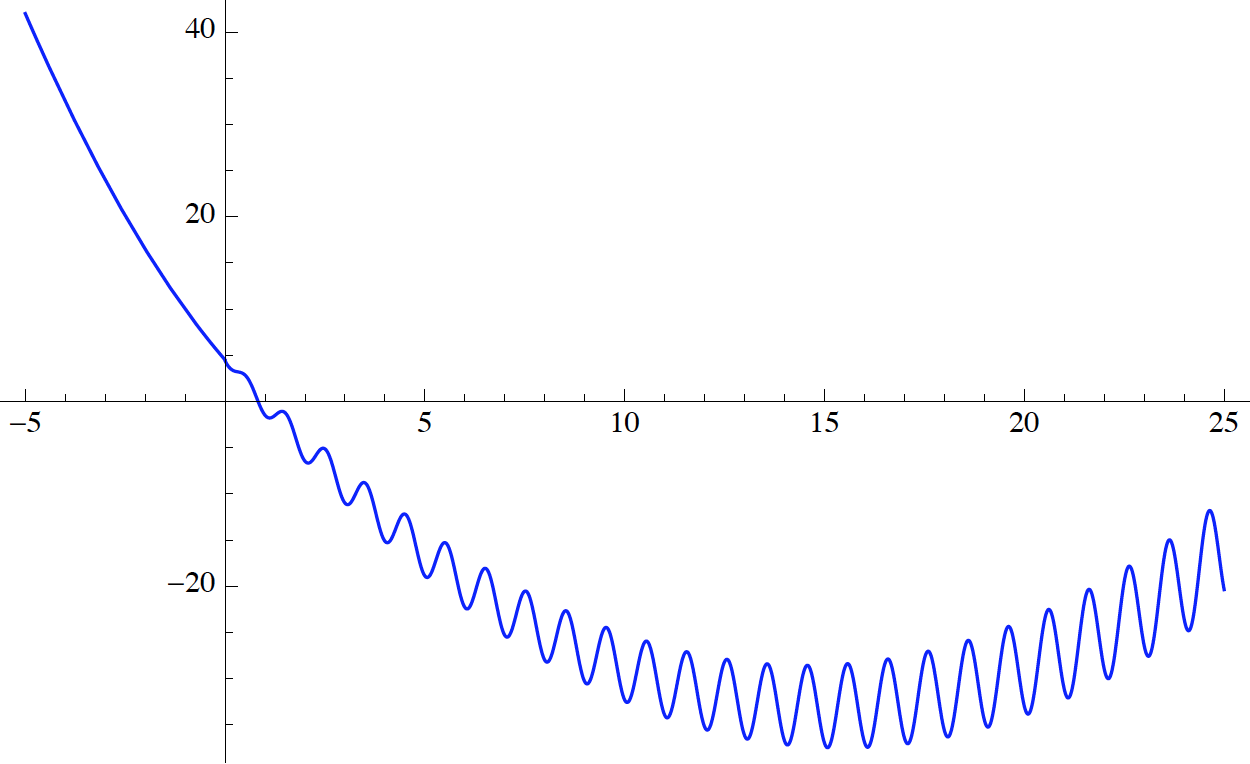}
 \put (22,57) {$V / M^4$} \put (88,35) {$\tilde{\phi} / \frac{M^2}{g}$} \put (80,57) {$\lambda=0.75$} \put (80,48) {$\kappa=1.0 M^3$}
\end{overpic}
 }
 \ \hspace{0.5mm} \hspace{5mm} \
 \subfloat[][]{
 \begin{overpic}[width=0.46\textwidth]{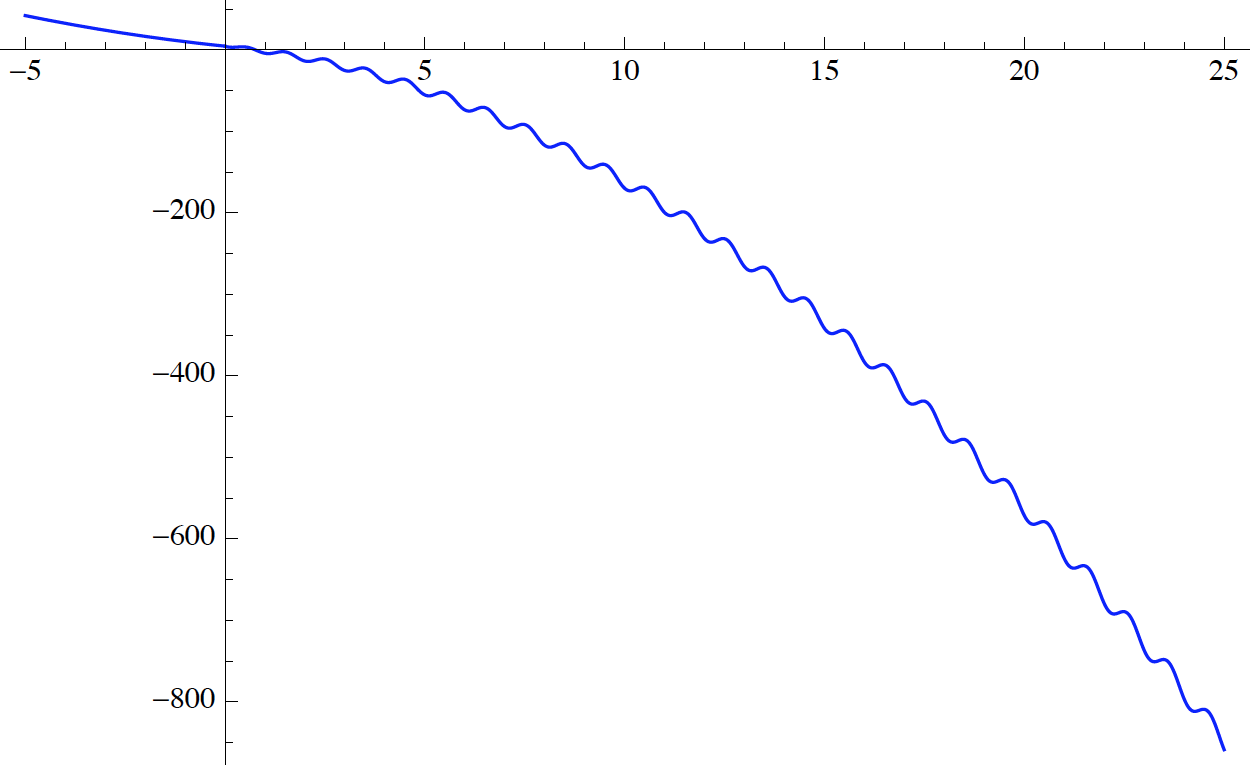}
 \put (22,2) {$V / M^4$} \put (88,48) {$\tilde{\phi} / \frac{M^2}{g}$} \put (22,37) {$\lambda=0.15$} \put (22,28) {$\kappa=1.0 M^3$}
\end{overpic}
 }
\hspace{5mm}
 \subfloat[][]{
\begin{overpic}[width=0.46\textwidth]{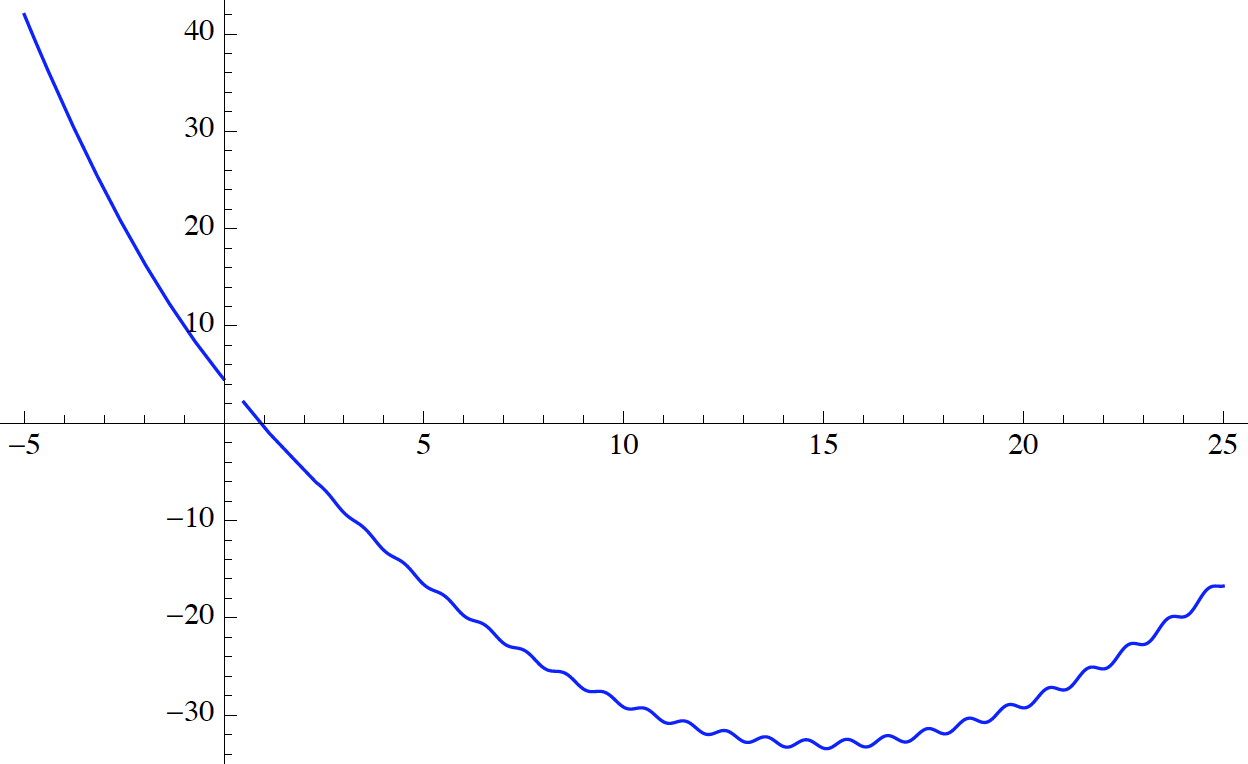}
 \put (22,57) {$V / M^4$} \put (88,33) {$\tilde{\phi} / \frac{M^2}{g}$} \put (80,57) {$\lambda=0.75$} \put (80,48) {$\kappa=0.1 M^3$}
\end{overpic}
 }
 \ \hspace{0.5mm} \hspace{5mm} \
 \subfloat[][]{
 \begin{overpic}[width=0.46\textwidth]{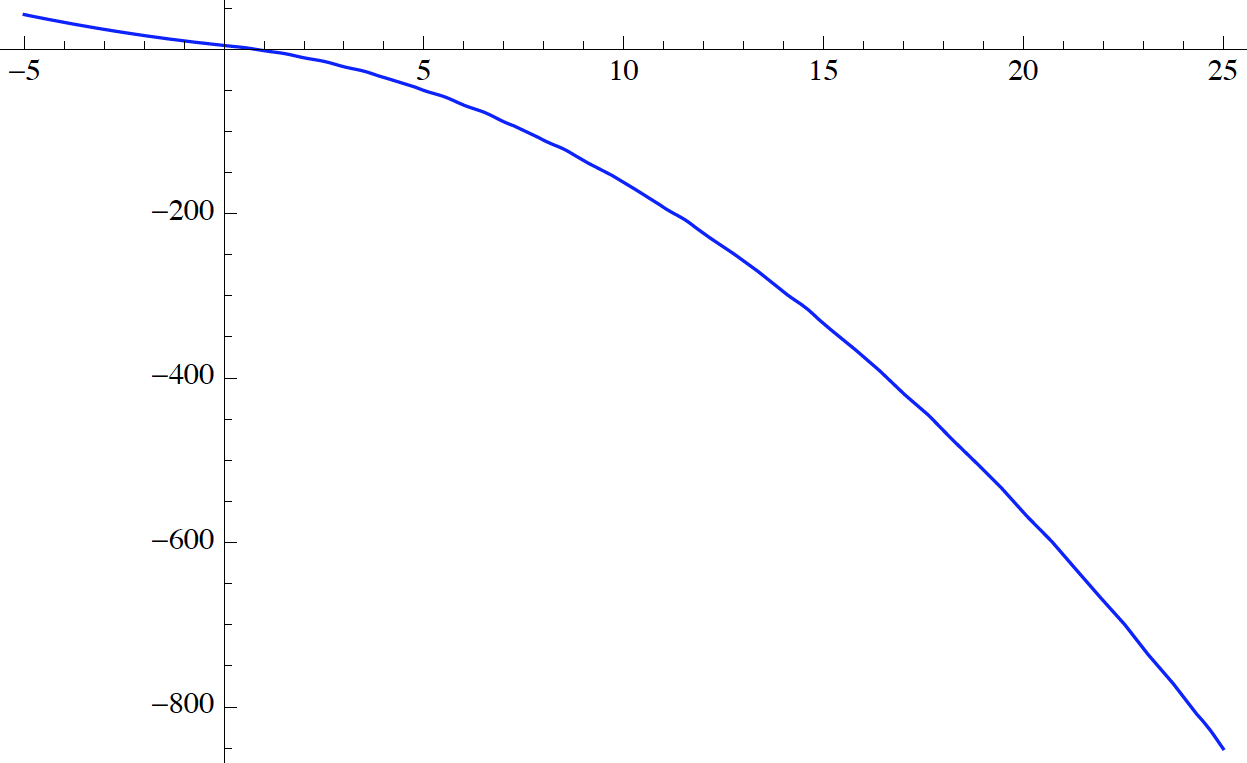}
 \put (22,2) {$V / M^4$} \put (88,48) {$\tilde{\phi} / \frac{M^2}{g}$} \put (22,37) {$\lambda=0.15$} \put (22,28) {$\kappa=0.1 M^3$}
\end{overpic}
 }
\caption{Plot of the effective axion potential for $c_1=4.0$, $c_2=1.0$, $M=0.001 M_{pl}$, $g=0.002 M$ and $f=80 M$.}
\label{fig:stability}
\end{figure}
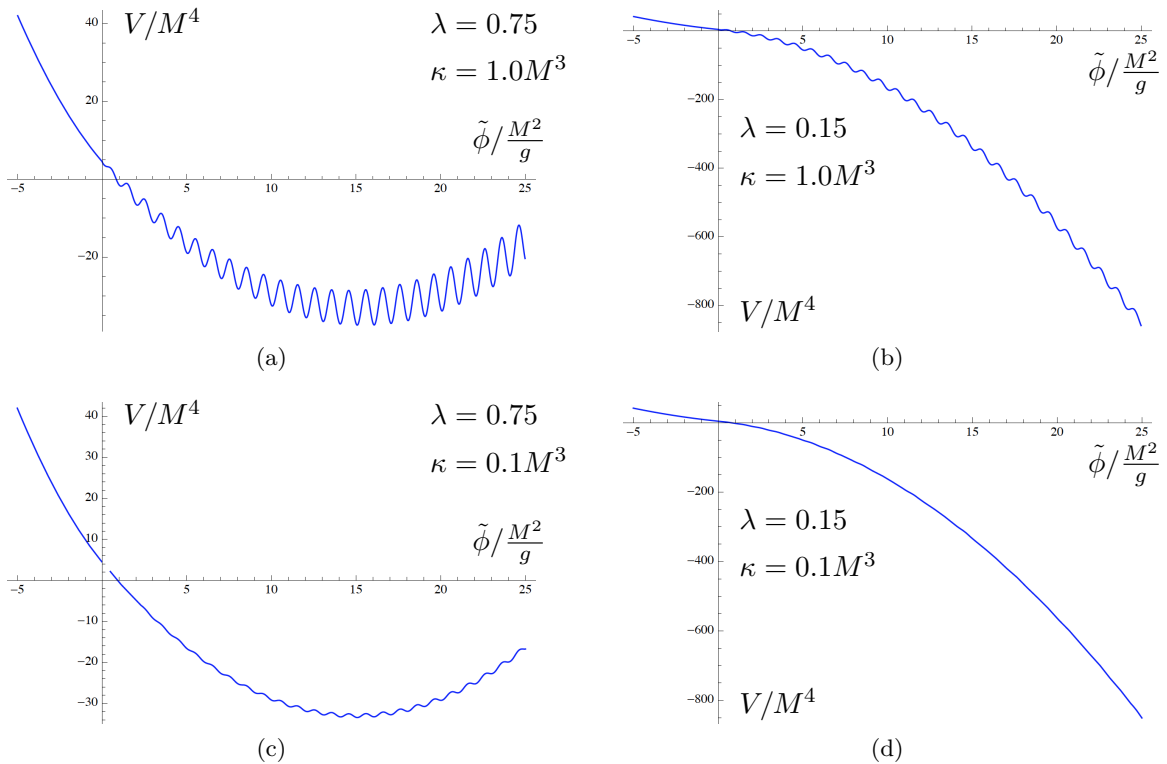

\noindent Thus, we obtain the effective potential for the axion field,
\begin{eqnarray}
\label{eq:singleV}
V(\tilde{\phi})
&=&\bigg\{\begin{array}{ll}
\frac{1}{2}g^2\left(c_{2}-\frac{1}{2\lambda}\right)\tilde{\phi}^2-gM^2\left(c_{1}+c_{2}\right)\tilde{\phi}-\kappa\frac{\sqrt{g\tilde{\phi}}}{\sqrt{\lambda}}\cos\!\left(\frac{\tilde{\phi}+\phi_{c}}{f}\right)+const.& \tilde{\phi}>0 \ , \qquad
\\
\frac{1}{2}g^2c_{2}\tilde{\phi}^2-gM^2\left(c_{1}+c_{2}\right)\tilde{\phi} +const.& \tilde{\phi}<0 \ . \qquad
\end{array}
\end{eqnarray}
Different values of the constants then result in different behaviors of the potential.  For $\tilde{\phi} <0$: 
\begin{itemize}
\item With $c_2 <0$, the potential has an instability as $\tilde{\phi} \rightarrow -\infty$. This is problematic, as the relaxation mechanism requires the field $\tilde{\phi}$ to start at a negative value. Instead of rolling towards larger field values, the axionic field exhibits a runaway behavior towards $\tilde{\phi} \rightarrow -\infty$. This is avoided if we take $c_2 >0$; 
\item For $c_1+c_2 <0$ the potential has a minimum at $\tilde{\phi}_{min} <0$. This is again undesirable. The relaxation mechanism requires $\tilde{\phi}$ to roll from a negative initial value to at least $\tilde{\phi}=0$. In particular, $\tilde{\phi}$ should not be trapped in a minimum, $\tilde{\phi}_{min} <0$. Thus, we require $c_1+c_2 > 0$.
\end{itemize}
Let us also note a couple of features of the polynomial part of the potential (i.e.~ignoring the cos-term) for $\tilde{\phi}>0$. 
\begin{itemize}
\item Now, the factor in front of the quadratic term is different. This arises from the fact that the energy is lowered when the Higgs acquires a non-vanishing vev. 
For $c_{2}<\frac{1}{2\lambda}$,\footnote{If $c_{2}$ is generated by a perturbative quantum correction one would expect this to be the case.}
this leads to an unstable direction as $\tilde{\phi}\to \infty$.
This is not a fundamental problem, since it is desirable to end up in a metastable minimum. 
However, if the Hubble friction is insufficiently large, i.e.~the Hubble constant is too small, the field can never stabilize; 
\item In contrast, for $c_{2}>\frac{1}{2\lambda}$, the polynomial potential for $\tilde{\phi}$ has a global minimum, for some $\tilde{\phi} >0$, and there is no danger of runaway behavior.
\end{itemize}
This observation persists when we include the cosine term in the potential. For $c_{2} > \frac{1}{2\lambda}$, the envelope of the potential is convex in the region $\tilde{\phi}>0$, while for $c_{2}<\frac{1}{2\lambda}$ it is concave. This is shown in fig.~\ref{fig:stability}.\footnote{From the figure one also notices that, for field ranges of order $\sim \phi_{c}$, the quadratic term can never be fully neglected, since $$ g^2\phi^{2}_{c}\sim gM^2\phi_{c} \ . $$}

While the overall envelope of the potential is governed by the polynomial parts, the oscillatory part is crucial for the detailed dynamics of the fields $\tilde{\phi}$ and $h$ -- and hence for the success of the model. In particular, depending on the size of the oscillatory terms the behavior of the fields will be qualitatively different. We identify three regimes:
\begin{enumerate}
\item{} If $\kappa$ is sufficiently large compared to the slope generated by the rest of the potential, the effective potential starts exhibiting a series of pronounced oscillations as soon as $\tilde{\phi}=0$ is reached. This is shown in figures~\ref{fig:stability}(a) and (b). We note, however, that this is to some degree an artifact of the one-field approximation. As we will discuss below the full two field model exhibits minima even before $\tilde{\phi}=0$ is reached;
\item{} With $\kappa$ small compared to the (negative) slope of the potential, the axion potential first needs to develop, i.e.~the factor $\sqrt{g\tilde{\phi}/\lambda}$ in front of the cosine has to grow first before pronounced minima and maxima appear. This can be seen in fig.~\ref{fig:stability} (c).
However, typically, the oscillatory phase does not last forever: Since the quadratic term yields a linearly increasing slope, while the axion potential only grows like a square root of $\tilde{\phi}$, the oscillations die down and the potential smoothly increases or decreases;
\item{} Finally, for some parameter values, the oscillatory contribution is never big enough to produced explicit maxima and minima. Such an example is shown in fig.~\ref{fig:stability} (d).
\end{enumerate}

\subsection{Properties with respect to the hierarchy problem}
Having identified these regimes, we can investigate their properties with respect to the hierarchy problem. We will use the single-field approximation where suitable, but also turn to the full two-field dynamics. This will be important, as we will see the two-field dynamics can depart from the results obtained using the single-field approximation in phenomenologically interesting situations.

During our initial analysis, we will assume that $\tilde{\phi}$ is slowly rolling, i.e. that its velocity tracks the slope of the potential,
\begin{equation}
\dot{\tilde{\phi}}=\frac{V'(\tilde{\phi})}{3H},
\end{equation}
and leave the discussion of slow-roll conditions until later.
If this situation is (approximately) fulfilled, the evolution of $\tilde{\phi}$ stops as soon as the slope becomes positive.

\subsubsection*{Regime (1)}
In the previous section we found that the effective single field potential for $\tilde{\phi}$ exhibits different qualitative features depending on the parameters of the model. Here we begin by analyzing regime (1) as described in section \ref{sec:single}. 

Initially, let us remain in the setting of the single-field approximation. Here, the oscillatory contributions to the axion potential become immediately significant once $\tilde{\phi}>0$. In particular, the first oscillation of the axion potential will already create a maximum. Accordingly, the potential will have a positive slope somewhere in the region,
\begin{equation}
0\leq\tilde{\phi}\leq2\pi f,
\end{equation}
and the field stops in the corresponding minimum in this field range. The Higgs vev in this regime is then given by,
\begin{equation}
\label{eq:hvev1}
\langle h\rangle =\sqrt{\frac{g\tilde{\phi}}{\lambda}} \sim \sqrt{\frac{g f}{\lambda}} \ ,
\end{equation}
where the last approximation holds unless the position of the minimum is somewhat fine-tuned. For small enough $g \ll \lambda M^2 / f$, we arrive at a value for the Higgs vev which lies well below the cutoff $M$. A small $g$ is absolutely necessary for $\langle h\rangle \ll M$ as we typically have $f >M$. If this holds, this mechanism appears suitable as a candidate solution to the hierarchy problem.

\begin{figure}[t]
 \subfloat[][]{
\begin{overpic}[width=0.46\textwidth]{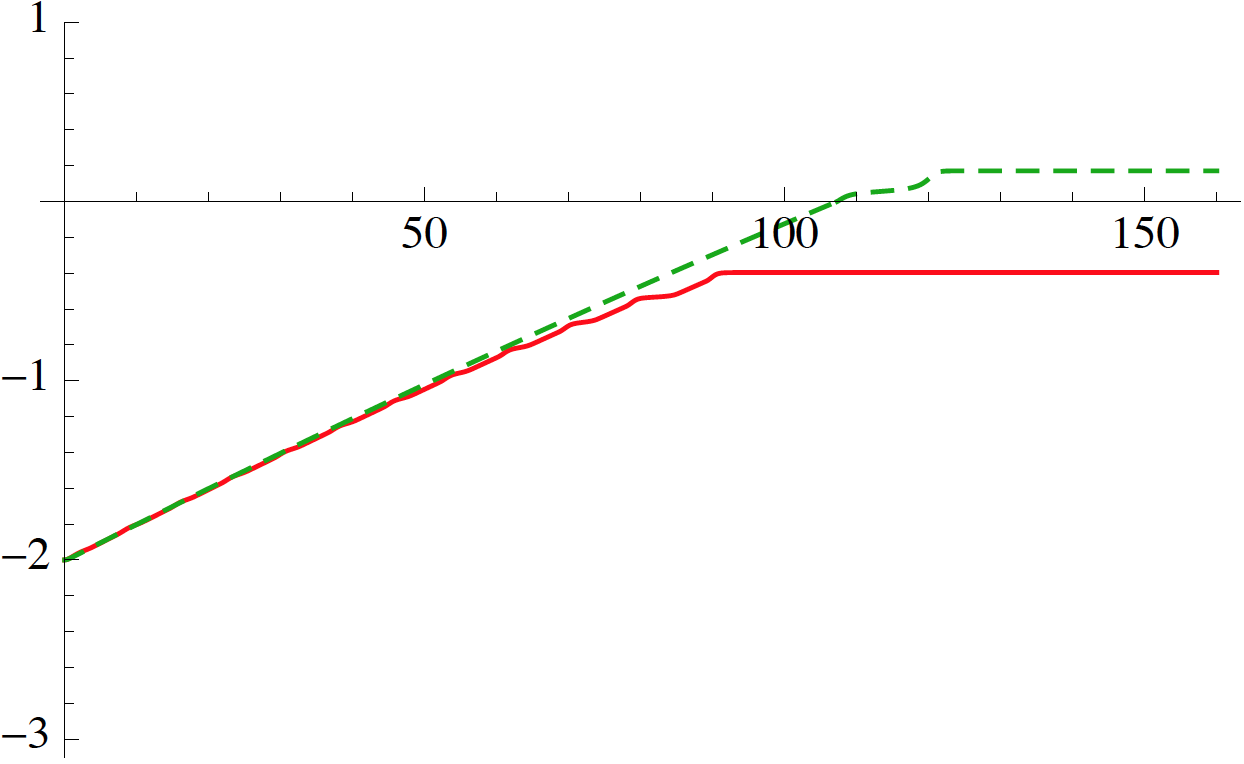}
 \put (92,55) {$t / H$} \put (10,5) {$\tilde{\phi} / \frac{M^2}{g}$}
\end{overpic}
 }
 \ \hspace{0.5mm} \hspace{5mm} \
 \subfloat[][]{
 \begin{overpic}[width=0.46\textwidth]{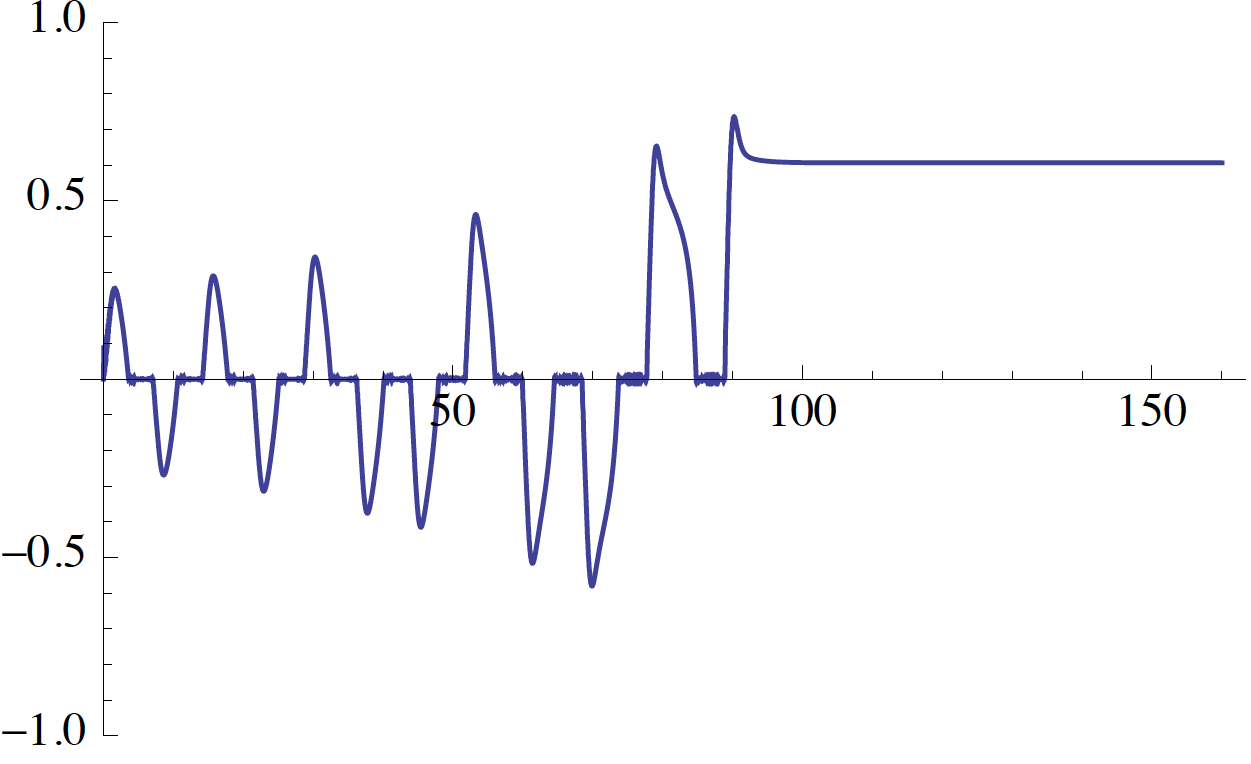}
 \put (12,58) {$h / M$} \put (88,35) {$t / H$}
\end{overpic}
 }
\caption{\emph{(a)}: Numerical result for the evolution of $\tilde{\phi}$ according to the full two-field model (red) and according to the single-field approximation (green, dashed). \emph{(b)}: evolution of $h$ for the full two-field model. Here we used $c_1=8.0$, $c_2=1.0$, $M=0.01 M_{pl}$, $g=0.0015 M$, $H=0.02 M$, $\lambda=0.5$, $\kappa=0.5 M^3$ and $f= 15 M$. The initial conditions are $\tilde{\phi}(0)= - 2 M^2/g$, $h(0)=10^{-6} M$ and $\dot{\tilde{\phi}}(0) = \dot{h}(0)=0$. For the two-field case the axion evolution stops before $\tilde{\phi}$ becomes positive, i.e.~before the Higgs mass becomes tachyonic.}
\label{Fig:tooearly}
\end{figure}

However, the above estimate for $\langle h\rangle$ was obtained in the single field approximation. Yet, in this regime this approximation does not faithfully represent the complete dynamics and we need to turn to the full two-field model. We will find the following difference between the two approaches. In the single-field approximation the Higgs is assumed to remain at $h=0$ for $\tilde{\phi} < 0$, but this will not be the case in the full two-field model. To be specific, once the axion starts rolling down its potential, the Higgs will not remain at exactly $h=0$. As the cosine-term in the potential \eqref{eq:tildeV} is linear in $h$, it shifts the true minimum of the potential for $h$ away from $h=0$. Thus, while $\tilde{\phi}$ is rolling down its potential, the Higgs will be displaced from $h=0$ (see e.g.~Fig.~\ref{Fig:tooearly}). 

This Higgs displacement is not included in the one-field discussion above and, as we will see, it leads to departures from the behavior of the single-field model. In particular, the amplitude of $h$ can become so large that the cosine-barriers for $\tilde{\phi}$, which grow like $\kappa |h|$, become large enough to stop the axion. In fact, this can occur in a regime where $\tilde{\phi} <0$, see Fig.~\ref{Fig:tooearly} (a), i.e.~in a regime where the Mexican-hat potential for the Higgs has not yet developed. 

To analyze this situation quantitatively, let us focus on the part of the potential \eqref{eq:tildeV} depending on $h$:
\be
\label{eq:Vh}
V \supset V_h \equiv - \frac{1}{2} g \tilde{\phi} h^2 + \frac{\lambda}{4} h^4 - \kappa |h| \cos \left(\frac{\tilde{\phi} + \phi_c}{f} \right) \ .
\ee
The Higgs field will track the minimum of this potential which, for both $h<0$ and $h \geq 0$, is given by
\be
\label{eq:falsemin}
\frac{\partial V}{\partial h} =0 \qquad \Rightarrow \qquad -g \tilde{\phi} |h| + \lambda |h|^3 - \kappa  \cos \left( \frac{\tilde{\phi}+\phi_c}{f} \right)=0 \ .
\ee
Note that for $\cos\!\left( \frac{\tilde{\phi}+\phi_c}{f}\right) <0$ the latter equation does not have a solution in the regime of interest, i.e.~for $\tilde{\phi}<0$. Nevertheless, one can check that the potential \eqref{eq:Vh} is minimized for $h=0$. On the other hand, for $\cos\!\left( \frac{\tilde{\phi}+\phi_c}{f}\right) >0$, there are minima with $h \neq 0$. In particular, the maximum displacement occurs when $\cos\left( \frac{\tilde{\phi}+\phi_c}{f}\right) =1$. Thus, we conclude that, for $\tilde{\phi} <0$, the Higgs field alternates between phases with $h=0$ and $h \neq0$ as shown in Fig.~\ref{Fig:tooearly} (b). 

We now wish to quantify when the displacements of $h$ become so large that they stop the evolution of $\tilde{\phi}$:  In the course of one period of $\cos\left( \frac{\tilde{\phi}+\phi_c}{f}\right)$, the Higgs field alternates between $h=0$ and a maximum displacement $h_{max}$. Furthermore, during one period, the contribution of the Higgs potential to the overall energy also changes. In particular, for $h=h_{max}$ the potential energy in the Higgs sector is lower than for $h=0$. We represent this by $\Delta V_h = V_h(h_{max}) - V_h(0) = V_h(h_{max})$. 

In addition, during one period of $\cos\!\left( \frac{\tilde{\phi}+\phi_c}{f}\right)$, the contribution of the polynomial part of the $\tilde{\phi}$-potential to the overall energy also changes. In fact, the potential energy is lowered as $\tilde{\phi}$ rolls towards larger field values. For small $|\tilde{\phi}| < M^2/g$, the polynomial part in \eqref{eq:tildeV} is well-approximated by the linear term. Thus, in the course of one period $\Delta \tilde{\phi} \sim f$, the energy changes by 
\be
\label{eq:Vpolearly}
\Delta V_{pol} \sim -(c_1+c_2) g M^2 f \ .
\ee

We now have all the necessary ingredients to determine whether the evolution will cease before reaching $\tilde{\phi}=0$. Let's take a value $\tilde{\phi}_0 <0$ where $\cos\!\left( \frac{\tilde{\phi}_0+\phi_c}{f}\right)=1$ and $h=h_{max}$. If $\tilde{\phi}$ continues to evolve, it will eventually reach a value where $\cos\!\left( \frac{\tilde{\phi}+\phi_c}{f}\right)$ will become negative again, at which point the Higgs field will return to $h=0$. However, returning $h$ to $h=0$ comes with an energy cost of $|\Delta V_h|$. For this to be possible, the increase in energy has to be offset by $\Delta V_{pol}$. As long as $|\Delta V_{pol}| > |\Delta V_h|$, it is then energetically favorable for $\tilde{\phi}$ to continue rolling. For $|\Delta V_{pol}| < |\Delta V_h|$ though, any further evolution of $\tilde{\phi}$ is impossible. Instead, the Higgs field will remain at its minimum, at $h = h_{max}$, and the axion will remain trapped at $\tilde{\phi} = \tilde{\phi}_0$. 

Thus, the evolution of $\tilde{\phi}$ will stop prematurely when
\be
|\Delta V_h| \sim |\Delta V_{pol}| \sim (c_1+c_2) g M^2 f \ .
\ee
All that remains is to find an expression for $\Delta V_h$. To proceed, we distinguish between two scenarios: the Higgs potential is dominated by its mass term and; the quartic term dominates.
\begin{itemize}
  \item \emph{Mass term dominates:} \textit{i.e.} $-\frac{1}{2} g \tilde{\phi} h^2 \gg \frac{\lambda}{4} h^4$: \newline
Setting $\cos\!\left( \frac{\tilde{\phi}_0+\phi_c}{f}\right) =1$ in  \eqref{eq:falsemin}, allows us to determine $h_{max}$. Ignoring the $\lambda$-dependent term, we find
\be
|h_{max}| = -\frac{\kappa}{g \tilde{\phi}_0} \ .
\ee
Substituting this into the Higgs-dependent potential \eqref{eq:Vh}, and ignoring the quartic term, one obtains
\be
\Delta V_h \approx - \frac{\kappa^2}{2 g \tilde{\phi}_0} + \frac{\kappa^2}{g \tilde{\phi}_0} = \frac{\kappa^2}{g \tilde{\phi}_0} = - \frac{\kappa}{2} |h_{max}| \ .
\ee
Comparing $|\Delta V_h| \sim |\Delta V_{pol}|$, results in an estimate for the Higgs vev. Dropping the absolute value sign and the subscript, we arrive at
\be
\nonumber \langle h \rangle \sim (c_1+c_2) \frac{g f M^2}{\kappa} \ .
\ee
We can also estimate for which parameter choices we expect to get stuck prematurely. This, in fact, occurs when
\be
|\Delta V_h| \gtrsim |\Delta V_{pol}| \qquad \Leftrightarrow \qquad \frac{\kappa^2}{g |\tilde{\phi}_0|} \gtrsim (c_1+c_2) g f M^2 \ .
\ee
While this relation is not particularly illuminating, one can draw one interesting conclusion from it: that the effect of stopping prematurely is the likeliest outcome once $g$ is chosen to be sufficiently small.

\item \emph{Quartic term dominates:} \textit{i.e.} $-\frac{1}{2} g \tilde{\phi} h^2 \ll \frac{\lambda}{4} h^4$: \newline
We repeat the above analysis but this time ignore the Higgs mass term. Setting $\cos\!\left( \frac{\tilde{\phi}_0+\phi_c}{f}\right) =1$ in \eqref{eq:falsemin} we now have
\be
|h_{max}| = \frac{\kappa^{1/3}}{\lambda^{1/3}} \ .
\ee
From \eqref{eq:Vh}, we then obtain
\be
\Delta V_h \approx \frac{\lambda}{4} \frac{\kappa^{4/3}}{\lambda^{4/3}} - \kappa \frac{\kappa^{1/3}}{\lambda^{1/3}} = - \frac{3}{4} \frac{\kappa^{4/3}}{\lambda^{1/3}} = - \frac{3}{4}\kappa |h_{max}| \ .
\ee
Comparing $|\Delta V_h| \sim |\Delta V_{pol}|$, we again arrive at the same expression for the Higgs vev (up to numerical factors which can be ignored):
\be
\nonumber \langle h \rangle \sim (c_1+c_2) \frac{g f M^2}{\kappa} \ .
\ee
As before, we can also check when one expects to be trapped for $\tilde{\phi} <0$:
\be
|\Delta V_h| \gtrsim |\Delta V_{pol}| \qquad \Leftrightarrow \qquad \frac{\kappa^{4/3}}{\lambda^{1/3}} \gtrsim (c_1+c_2) g f M^2 \ .
\ee
Again, we expect this regime to occur for sufficiently small $g$.
\end{itemize}

\noindent An important finding of the above discussion is that, if the axion evolution ceases for $\tilde{\phi} <0$, we expect the Higgs vev to depend on the model parameters as
\be
\label{eq:hvev2}
\langle h \rangle \sim (c_1+c_2) \frac{g f M^2}{\kappa} \ .
\ee
Note the difference from our expectation using the single-field approximation \eqref{eq:hvev1}. The expression~\eqref{eq:hvev2} is identical to that suggested in~\cite{Graham:2015cka}, as can be inferred from their Eq.~(7) with the identification $\kappa=\Lambda^4/v$ and $v=246\,{\rm GeV}$ (i.e.~$v$ is fixed and is not the dynamically generated Higgs vev).

However, recall that for $\langle \tilde{\phi} \rangle <0$, the potential for the Higgs potential is not of Mexican-hat form. In fact, it would take the form
\be
V_h \sim  \frac{m^2}{2} h^2 + \frac{\lambda}{4} h^4 - \kappa |h| \ ,
\ee
with $m^2= g |\langle\tilde{\phi}\rangle|$. In this case the source of electroweak symmetry breaking is the linear term $\sim \kappa |h|$ rather than a tachyonic mass term. Thus, the vacuum for the Higgs differs from the vacuum in the Standard Model, which can have observable consequences.
In the QCD model this vev is obviously too small for realistic phenomenology\footnote{It is also not clear whether the structure of the oscillating term $\sim \kappa |h|\cos(\phi/f)$ persists for such small values for $h$.}.
In non-QCD models it is conceivable that a realistic Higgs vev is attained in this phase. It would be interesting to deduce any new experimental signatures for such a different vacuum. A particularly interesting observable would then be the Higgs self-coupling~\cite{Djouadi:1999gv,Baur:2002rb,Baur:2002qd,Baur:2003gpa,Baur:2003gp,Dolan:2012rv,Baglio:2012np,Barr:2014sga}. We leave a proper study of this for future work.

\subsubsection*{Regime (2)}
We now turn to regime (2), as described in section \ref{sec:single}. In the single-field approximation, this corresponds to a choice of parameters such that the oscillatory potential does not immediately lead to a series of prominent maxima and minima once $\tilde{\phi}$ crosses zero: the size of the oscillatory `bumps' has to grow before pronounced maxima and minima appear. 

Our goal, as before, is to derive an expression for the final value of the axion $\tilde{\phi}$ once it is trapped. The corresponding Higgs vev then follows from $| \langle h \rangle |\sim \sqrt{g \tilde{\phi} / \lambda}$. Obviously, the axion can only get trapped if we have pronounced minima. Thus, we begin our analysis by determining the range in $\tilde{\phi}$-space where maxima and minima occur. This occurs when the slope of the oscillatory part of the $\tilde{\phi}$-potential \eqref{eq:singleV} dominates that of the polynomial part. If we assume that $\tilde{\phi}$ is slowly rolling, it will then be trapped by the first minimum it encounters. 

For the polynomial part, we have, for $\tilde{\phi}>0$,
\begin{equation}
V'_{\rm polynomial}=g^2\left(c_{2}-\frac{1}{2\lambda}\right)\tilde{\phi}-gM^2(c_{1}+c_{2})=g^2d_{2}\tilde{\phi}-gM^2d_{1},
\end{equation}
where 
\begin{equation}
d_{2}=\left(c_{2}-\frac{1}{2\lambda}\right),\qquad d_{1}=(c_{1}+c_{2}).
\end{equation}

As explained in section \ref{sec:single}, we require $d_1>0$, while $d_2$ can be positive or negative. The oscillatory part of the potential contains two terms,
\begin{equation}
V'_{\rm osc}=\frac{\kappa\sqrt{g}}{\sqrt{\lambda}}\left[\frac{\sqrt{\tilde{\phi}}}{f}\sin\left(\frac{\tilde{\phi}+\phi_{c}}{f}\right)
-\frac{1}{2\sqrt{\tilde{\phi}}}\cos\!\left(\frac{\tilde{\phi}+\phi_{c}}{f}\right)\right].
\end{equation}
For $\tilde{\phi}\gg f$, we then have,
\begin{equation}
|V'_{\rm osc}|\sim \frac{\kappa\sqrt{g}}{\sqrt{\lambda}}\frac{\sqrt{\tilde{\phi}}}{f} \ .
\end{equation}

Minima and maxima are present when the slope from the axion potential exceeds that of the polynomial part, \textit{i.e.}
\begin{equation}
\label{eq:reg2minmax}
|V'_{\rm osc}| > |V'_{\rm polynomial}| \ .
\end{equation}
We can solve this quadratic equation to determine the range of $\tilde{\phi}$ where maxima and minima occur. Using $| \langle h \rangle |= \sqrt{g \tilde{\phi} / \lambda}$ we can then also determine the corresponding range of $\langle h\rangle$. 

For $d_2<0$, we obtain:
\begin{equation}
\label{eq:rangel0}
-\sqrt{\frac{\kappa^2}{4 |d_2|^2 f^2 g^2 \lambda^2}-\frac{M^2d_{1}}{\lambda |d_{2}|}} + \frac{\kappa}{2 |d_2| fg \lambda} < | \langle h\rangle | < \sqrt{\frac{\kappa^2}{4 |d_2|^2 f^2 g^2 \lambda^2}-\frac{M^2d_{1}}{\lambda |d_{2}|}} + \frac{\kappa}{2 |d_2| fg \lambda} \ ,
\end{equation}
while for $d_2 > 0$ we find:
\begin{equation}
\label{eq:rangeg0}
\sqrt{\frac{\kappa^2}{4 d^{2}_{2} f^2 g^2 \lambda^2}+\frac{M^2d_{1}}{\lambda d_{2}}} - \frac{\kappa}{2d_2 fg \lambda} < | \langle h\rangle | < \sqrt{\frac{\kappa^2}{4 d^{2}_{2} f^2 g^2 \lambda^2}+\frac{M^2d_{1}}{\lambda d_{2}}} + \frac{\kappa}{2d_2 fg \lambda} \ .
\end{equation}
As long as $\tilde{\phi}$ is slowly rolling, it will get trapped by the first minimum it encounters. The corresponding vev for $h$ will then be given by the lower end of the ranges \eqref{eq:rangel0} and \eqref{eq:rangeg0}. However, the above formulae are not very illuminating. To extract some meaning let us distinguish three scenarios:
\begin{itemize}
\item Case 1: $\frac{M^2d_{1}}{\lambda |d_{2}|} \ll \frac{\kappa^2}{4 |d_{2}|^{2} f^2 g^2 \lambda^2}$\newline
This situation can occur if, for example, $g$ is sufficiently small. In this case, the lower threshold for $\langle h \rangle$ approaches
\be
\langle h \rangle \sim d_1 \frac{fgM^2}{\kappa}
\ee
for both $d_2>0$ and $d_2<0$. This is identical to the result \eqref{eq:hvev2} we found in regime (1). For sufficiently small $g$ the Higgs vev can be well below the cutoff $M$ as required to solve the hierarchy problem.;
\item Case 2: $\frac{M^2d_{1}}{\lambda |d_{2}|} \sim \frac{\kappa^2}{4 |d_{2}|^{2} f^2 g^2 \lambda^2}$\newline
For parameter choices leading to this case one finds that 
\be
\langle h \rangle \sim \sqrt{\frac{d_1}{|d_2| \lambda}} M \sim M \ ,
\ee
for both $d_2>0$ and $d_2<0$.
Clearly, this does not address the hierarchy problem;
\item Case 3: $\frac{M^2d_{1}}{\lambda |d_{2}|} \gg \frac{\kappa^2}{4 |d_{2}|^{2} f^2 g^2 \lambda^2}$\newline
This situation arises when the oscillatory contribution to the $\tilde{\phi}$-potential is always subleading. This, by definition, corresponds to what we termed regime (3) in section \ref{sec:single}. We will discuss this case separately.
\end{itemize}

\begin{figure}[t]
 \subfloat[][]{
\begin{overpic}[width=0.46\textwidth]{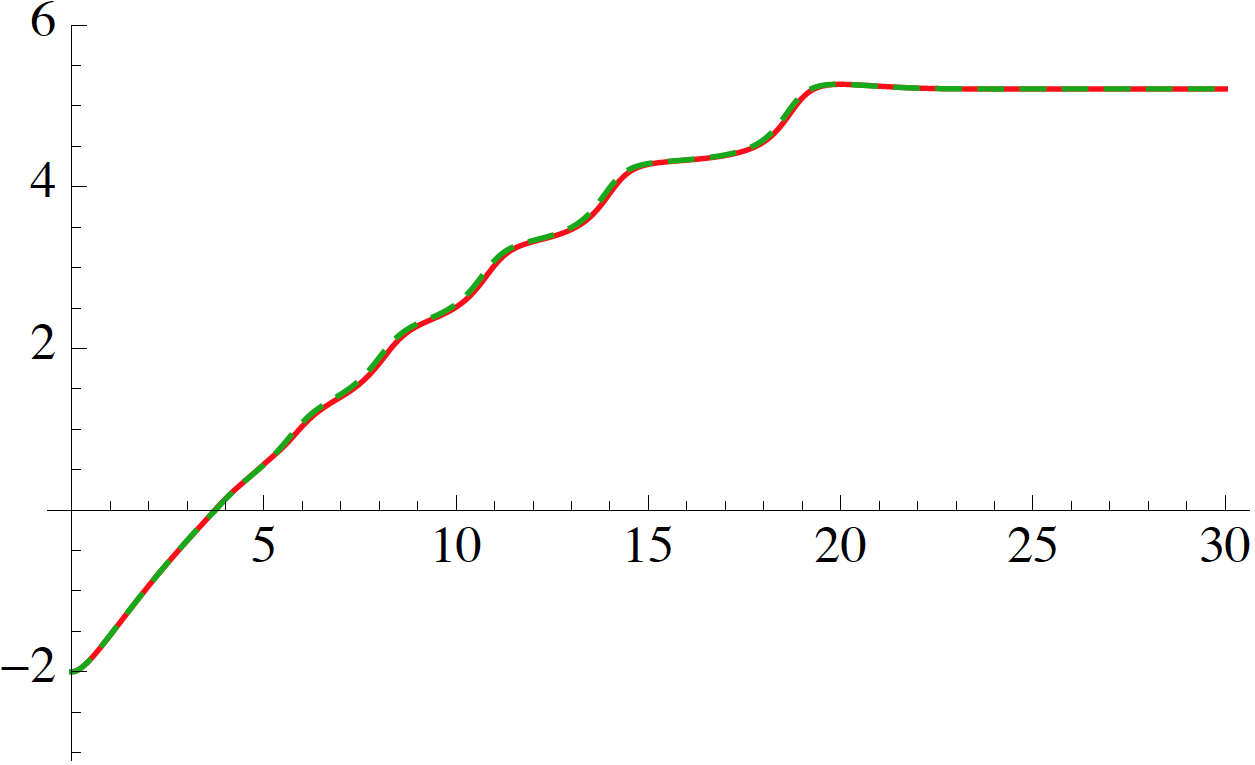}
 \put (92,25) {$t / H$} \put (10,58) {$\tilde{\phi} / \frac{M^2}{g}$}
\end{overpic}
 }
 \ \hspace{0.5mm} \hspace{5mm} \
 \subfloat[][]{
 \begin{overpic}[width=0.46\textwidth]{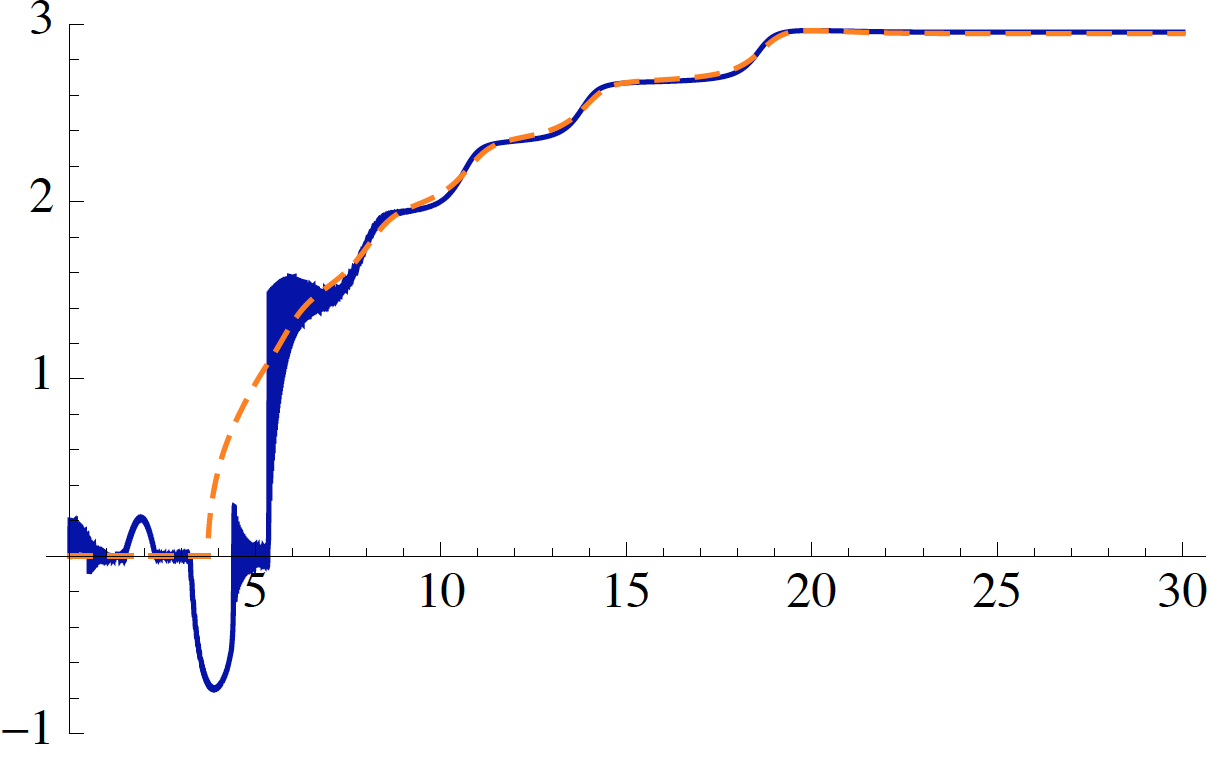}
 \put (10,58) {$h / M$} \put (88,21) {$t / H$}
\end{overpic}
 }
\caption{\emph{(a)}: Numerical result for the evolution of $\tilde{\phi}$ according to the full two-field model (red) and according to the single-field approximation (green, dashed). \emph{(b)}: evolution of $h$ according to the full two-field model (blue) and according to the single-field approximation (orange, dashed). Here we used $c_1=3.6$, $c_2=1.0$, $M=0.01 M_{pl}$, $g=0.002 M$, $H=0.036 M$, $\lambda=0.6$, $\kappa=0.22 M^3$ and $f= 80 M$. The initial conditions are $\tilde{\phi}(0)= - 2 M^2/g$, $h(0)=10^{-6} M$ and $\dot{\tilde{\phi}}(0) = \dot{h}(0)=0$. The single-field approximation matches the full two-field dynamics very closely.}
\label{fig:reg2}
\end{figure}

In summary, for parameter choices leading to case 1 above, the relaxation mechanism offers a possible solution to the hierarchy problem. We then find the following conditions on the model parameters:
\begin{align}
\label{eq:reg2cond1}
\frac{M^2d_{1}}{\lambda |d_{2}|} & \ll \frac{\kappa^2}{4 |d_{2}|^{2} f^2 g^2 \lambda^2} \qquad \textrm{(``case 1'')} \\ 
\label{eq:reg2cond2}
{\langle h \rangle} \sim d_1 \frac{fgM^2}{\kappa}  & \ll M \qquad \textrm{(``address hierarchy problem'')} \ .
\end{align}
There is one caveat: to be in regime (2) we require $\tilde{\phi} \gg f$. This was explicitly used when deriving \eqref{eq:rangel0} and \eqref{eq:rangeg0}. If this condition is not met and and instead $\langle \tilde{\phi} \rangle \sim f$ we are back to regime (1) and the results from this section are not valid\footnote{Recall that in regime (1) the relaxation mechanism is prone to trapping $\tilde{\phi}$ ``prematurely'' with Higgs stabilized in a non-standard vacuum where the Higgs vev is induced by a source term.}. Hence, we also need to ensure that
\be
\label{eq:reg2cond3}
\langle \tilde{\phi} \rangle = \frac{\lambda}{g} {\langle h \rangle}^2 = d_1^2 \lambda \frac{f^2 g M^4}{\kappa^2} \gg f \qquad \textrm{(``regime (2)'')} \ .
\ee
These three conditions can be rewritten as follows. To make the discussion more transparent, note that generically $d_1 \sim d_2 \sim \mathcal{O}(1)$. Furthermore, $\lambda$ is determined by experiment and for the present purpose we can set $\lambda \sim \mathcal{O}(1)$. Then the three conditions \eqref{eq:reg2cond1}--\eqref{eq:reg2cond3} imply 
\be
\label{eq:kappabound}
{\left(\frac{fg}{M^2} \right)}^2 \ll {\left( \frac{\kappa}{M^3} \right)}^2 \ll \frac{fg}{M^2} \ ,
\ee
i.e.~$\kappa$ is bounded from above and below. It has to be large enough to obtain a hierarchically small Higgs vev but small enough to avoid regime (1).

Before moving on to regime (3), let us compare our findings of the single-field approximation with the full two-field dynamics. In figure \ref{fig:reg2} we display the evolution of $\tilde{\phi}$ and $h$ for both the single-field approximation and the two-field model. The parameter choice corresponds to Case 2 of the above list, i.e.~$\frac{M^2d_{1}}{\lambda |d_{2}|} \sim \frac{\kappa^2}{4 |d_{2}|^{2} f^2 g^2 \lambda^2}$. Overall, we find very good agreement between the single-field approximation and the full two-field dynamics. For $\tilde{\phi} >0$, the Higgs field closely follows its instantaneous minimum $h=\sqrt{g \tilde{\phi}/\lambda}$ as assumed in the single-field approximation. The Higgs vev is given by $\langle h \rangle \approx 3M$, which is consistent with the above analysis. Thus, for the given parameter choice, there is no hierarchy between $\langle h \rangle$ and $M$.

\subsubsection*{Regime (3)}
In regime (3), as described in section \ref{sec:single}, the oscillatory contribution to the effective $\tilde{\phi}$-potential is never large enough to produce maxima and minima. Clearly, the relaxation mechanism will not work and we cannot solve the hierarchy problem. Nevertheless, for completeness, let us briefly record the results using the single-field approximation.

In regime (3) we can ignore the oscillatory contribution to the potential and focus on the polynomial part.
For $d_2>0$, the polynomial $\tilde{\phi}$-potential has a minimum at 
\be
\tilde{\phi} = \frac{d_1}{d_2} \frac{M^2}{g} \ .
\ee
The corresponding Higgs vev is given by
\be
\langle h \rangle \sim \sqrt{\frac{d_1}{d_2 \lambda}} M \sim M \ ,
\ee
which does not solve the hierarchy problem. 

For $d_2<0$, the polynomial $\tilde{\phi}$-potential does not have any minima. As the oscillatory contribution is too small to produce minima the field $\tilde{\phi}$, and hence $h$, do not get stabilized and diverge.

\subsubsection*{Summary}
A relaxation mechanism employing a two-field potential of the form \eqref{eq:V} can indeed successfully produce a Higgs vev smaller than the cutoff. However, there are pitfalls. For certain parameter choices the axion can get trapped too early and the Higgs is stabilized in a non-standard vacuum. For other parameter choices there is no hierarchy between the Higgs vev and the UV cutoff. Overall, the success of a relaxation mechanism based on \eqref{eq:V} in solving the hierarchy problem is highly dependent on the model parameters. It is less robust against modifications of these parameters than one would initially expect.

\subsection{When do we have slow-roll?}
The conclusions above for solving the hierarchy problem hold only for the slow-roll regime.
Here, we briefly look at a sufficient condition for the axion field to be slowly rolling in the effective potential.

For a potential with a constant slope, $k$, the velocity, in field space, is,
\begin{equation}
\dot{\phi}=\dot{\phi}_{\infty}+(\dot{\phi}(t=0)-\dot{\phi}_{\infty})\exp(-3Ht),
\end{equation}
where 
\begin{equation}
\dot{\phi}_{\infty}=\frac{k}{3H}
\end{equation}
is the asymptotic velocity.

We see that the typical relaxation time for the field to return to its asymptotic velocity is,
\begin{equation}
t_{\rm relax}\sim \frac{1}{3H}.
\end{equation}

If we now look at a situation where the slope suddenly changes from $k$ to $0$, the field has travelled a distance,
\begin{equation}
\Delta\phi\sim \dot{\phi}(t=0)\frac{1}{3H}\sim \frac{k}{9H^2},
\end{equation}
until it has reached its new asymptotic velocity of 0.

For our purposes, we say a field is slow-rolling if it gets stuck in the first minimum, or in one of the first few minima.
This happens when the field relaxes to its new velocity in one oscillation,
\begin{equation}
\Delta\tilde{\phi}\lesssim f.
\end{equation}
In the vicinity of $\tilde{\phi}=0$, the polynomial potential for $\tilde{\phi}$ is approximately linear with slope
$d_{1}gM^2$. For slow roll we thus require 
\begin{equation}
\label{eq:slowroll3}
\frac{d_{1}gM^2}{9H^2}\lesssim f.
\end{equation}
Note that this condition is different from the slow-roll conditions \eqref{eq:slowroll1} and \eqref{eq:slowroll2} proposed in \cite{Graham:2015cka}.

\begin{figure}[t]
\begin{center}
\begin{overpic}[width=0.46\textwidth]{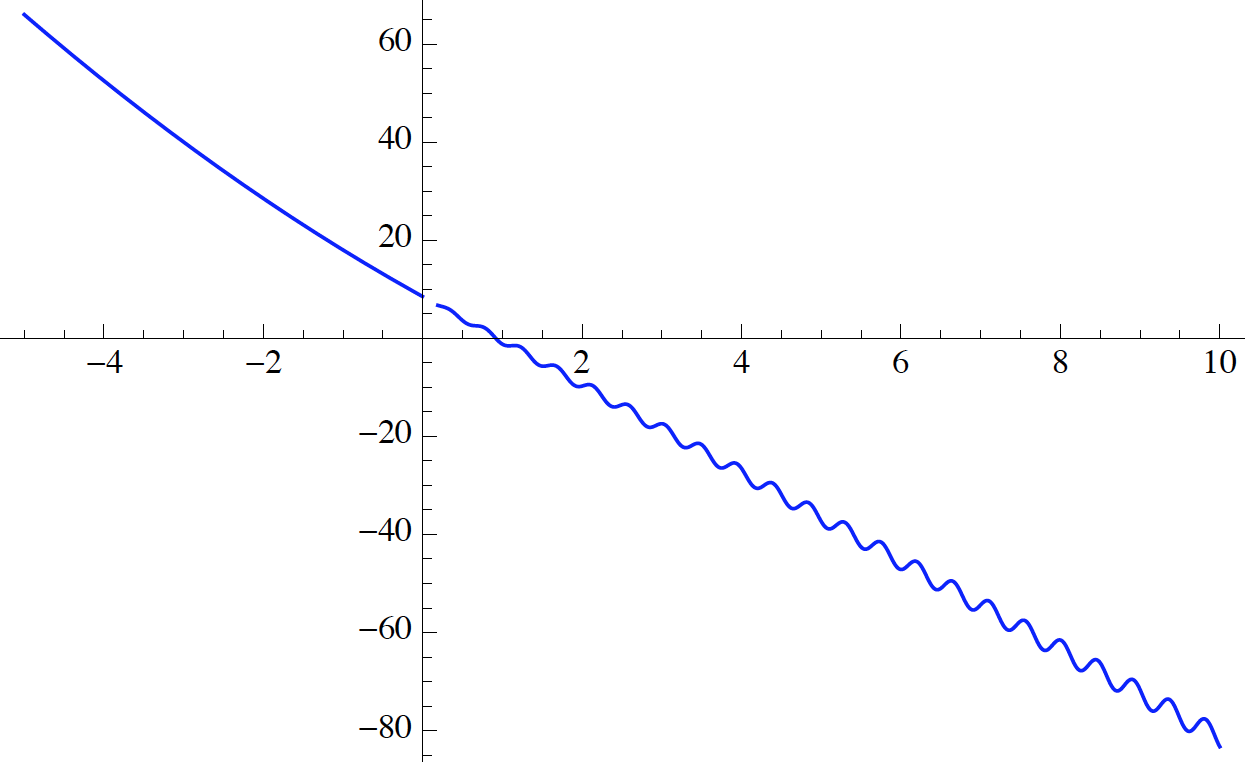}
 \put (38,57) {$V / M^4$} \put (88,25) {$\tilde{\phi} / \frac{M^2}{g}$}
\end{overpic}
\caption{Effective $\tilde{\phi}$-potential in the single-field approximation for $c_1=8.0$, $c_2=1.0$, $M=0.001 M_{pl}$, $g=0.0036 M$, $\lambda=0.5$, $\kappa=0.5 M^3$ and $f= 20 M$.}
\label{fig:fastrollpot}
\end{center}
\end{figure}

\begin{figure}[t]
 \subfloat[][]{
\begin{overpic}[width=0.46\textwidth]{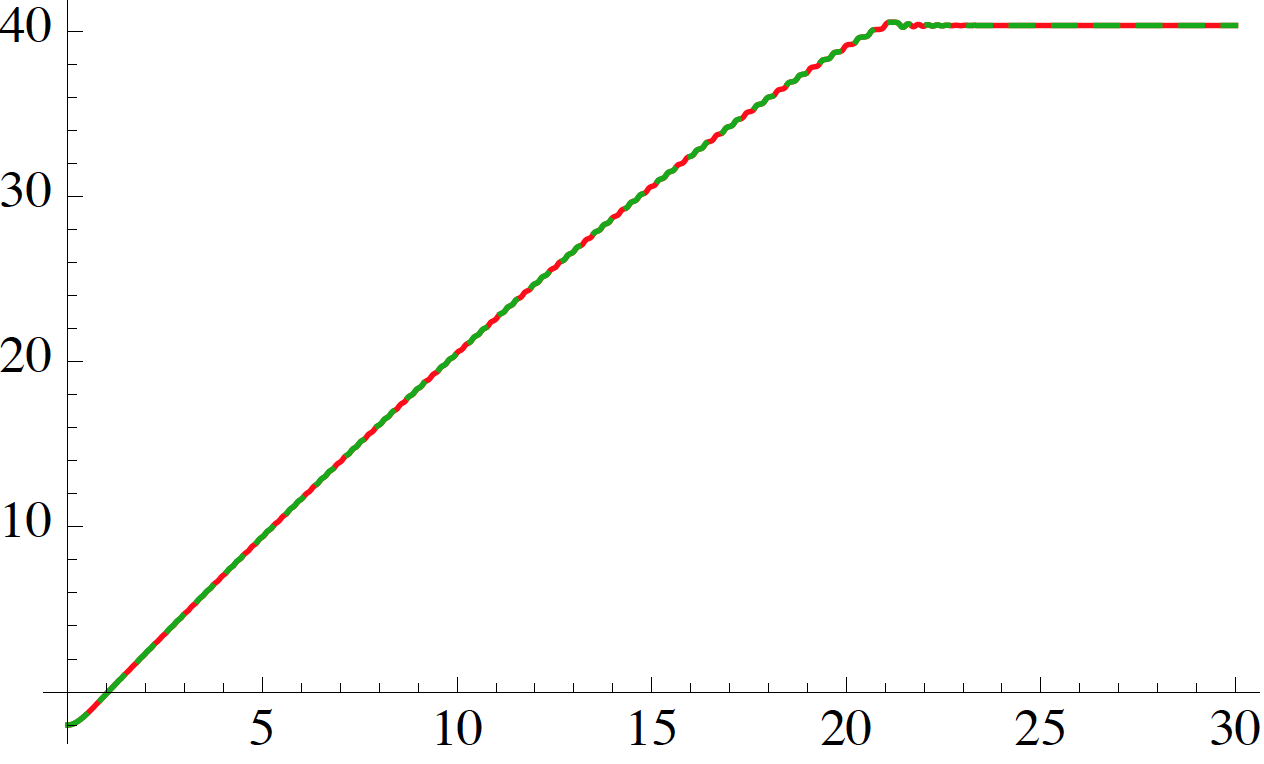}
 \put (92,10) {$t / H$} \put (10,61) {$\tilde{\phi} / \frac{M^2}{g}$}
\end{overpic}
 }
 \ \hspace{0.5mm} \hspace{5mm} \
 \subfloat[][]{
 \begin{overpic}[width=0.46\textwidth]{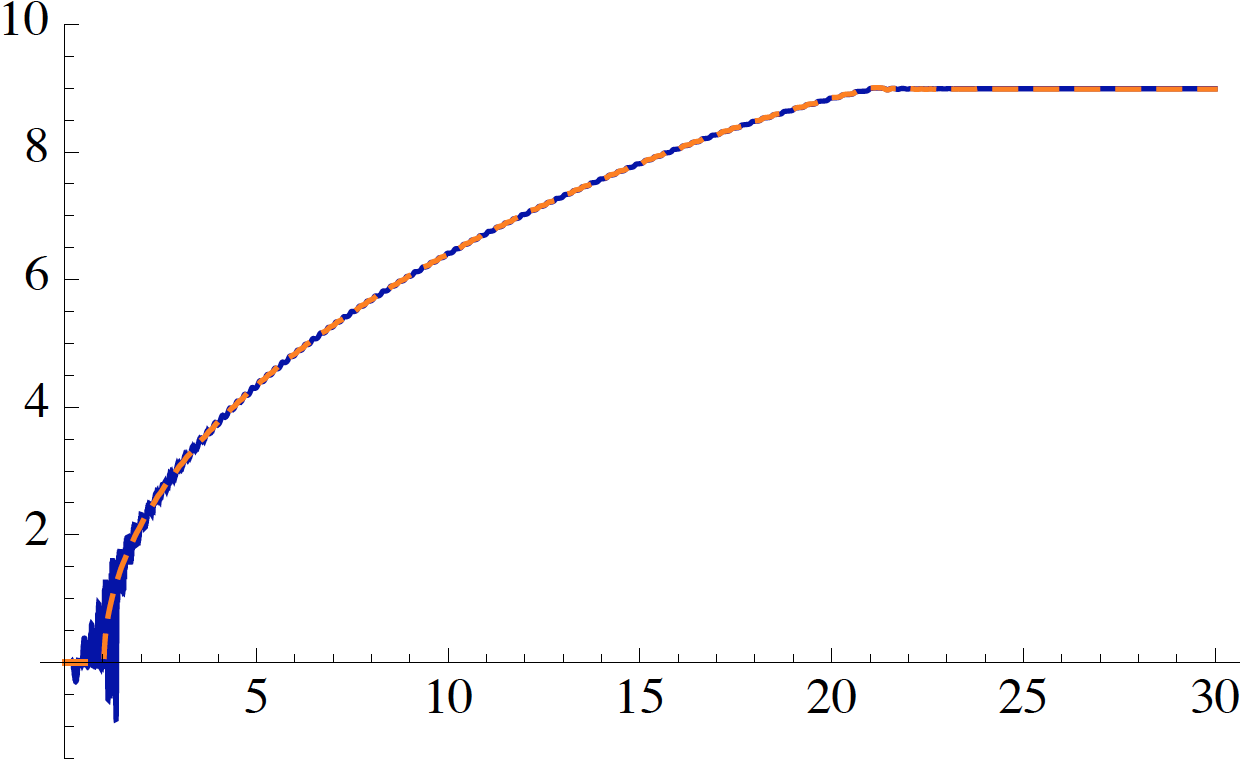}
 \put (10,61) {$h / M$} \put (92,13) {$t / H$}
\end{overpic}
 }
\caption{\emph{(a)}Numerical result for the evolution of $\tilde{\phi}$ according to the full two-field model (red) and according to the single-field approximation (green, dashed). \emph{(b)}: Evolution of $h$ according to the full two-field model (blue) and according to the single-field approximation (orange, dashed). Here we used $H=0.004 M$ and the model parameters of figure \ref{fig:fastrollpot}. The initial conditions are $\tilde{\phi}(0)= - 2 M^2/g$, $h(0)=10^{-6} M$ and $\dot{\tilde{\phi}}(0) = \dot{h}(0)=0$. }
\label{fig:fastroll}
\end{figure}

To exemplify the significance of the condition \eqref{eq:slowroll3}, let us consider an example. We choose parameters giving rise to the effective $\tilde{\phi}$-potential of figure \ref{fig:fastrollpot} which exhibits a series of pronounced maxima and minima once $\tilde{\phi} > 0$. If the axion was slowly rolling one would expect $\tilde{\phi}$ to get trapped by the first or one of the first few minima. 

We now choose $H$ such that the slow-roll conditions \eqref{eq:slowroll1} and \eqref{eq:slowroll2} of \cite{Graham:2015cka} are satisfied while \eqref{eq:slowroll3} is not. The resulting time evolution of $\tilde{\phi}$ and $h$ is displayed in figure \ref{fig:fastroll}. Rather than getting stuck in one of the first minima, the axion overshoots and only stabilizes at $\tilde{\phi} \sim 40 M^2/g$. Once $\tilde{\phi} >0$, the Higgs vev tracks its instantaneous minimum $\langle h \rangle = \sqrt{\tilde{g\phi} / \lambda}$ until the axion stops. The final vev of the Higgs is given by $\langle h \rangle \sim 9M$, which is well above the cutoff of the theory. 

This example illustrates that the conditions \eqref{eq:slowroll1} and \eqref{eq:slowroll2} of \cite{Graham:2015cka} are not sufficient to ensure that $\tilde{\phi}$ will get trapped in one of the first minima. However, as the final value of $\tilde{\phi}$ determines the Higgs vev through $\langle h \rangle = \sqrt{g \tilde{\phi} / \lambda}$, we do not want $\tilde{\phi}$ to be stabilized at too large a value if we wish to solve the hierarchy problem. To avoid dangerous overshooting we thus have to ensure that the field $\tilde{\phi}$ is sufficiently slowly rolling. This introduces the additional constraint \eqref{eq:slowroll3} on the model parameters.

\section{Discussion}
In the introduction we collected a set of principles for an effective field theory to be regarded as an improvement over the SM with regard to the electroweak hierarchy problem. In the light of our findings of section \ref{sec:dynamics} we will now discuss to what extent these principles are realized in the paradigm of ``Cosmological Relaxation'' introduced in \cite{Graham:2015cka}.

We found that for certain parameter choices the model of \cite{Graham:2015cka} is capable of producing a hierarchically small Higgs vev, which depends on the model parameters as
\be
\label{eq:hvevdisc}
v \equiv \langle h \rangle \sim \frac{fgM^2}{\kappa} \ .
\ee
Then $\langle h \rangle \ll M$ can be achieved by choosing $f$ and $g$ small and/or $\kappa$ large. However, the scale $f$ is expected to be $f \gtrsim M$ and thus cannot be chosen small.\footnote{For models employing the QCD axion this is necessary because constraints on the QCD axion require $f\gtrsim 10^{9}\,{\rm GeV}$ and so far no model has been found that allows for $M>10^{9}\,{\rm GeV}$. More generally one would like to have $f\sim M$ or at least $f\gtrsim M$ to avoid having to explain new hierarchies.} In addition, note that $\kappa$ is bounded (see \eqref{eq:kappabound}) and taking it too large would stabilize the Higgs in a non-standard vacuum.\footnote{Furthermore, if we take $\phi$ to be the QCD axion the parameters $f$ and $\kappa$ are determined by QCD and cannot be adjusted.} As a result, a small Higgs vev can only be achieved by taking $g$ to be sufficiently small. For the following discussion let us define $\epsilon \equiv g/M$ as the relevant dimensionless small parameter.

We now proceed to addressing the nine points raised in the introduction.

\subsubsection*{Ad 1) and 2)}
The mechanism of \cite{Graham:2015cka} requires at least one small parameter, $\epsilon$, to explain a hierarchically small Higgs vev. It thus exchanges a tuning of the Higgs mass parameter for the tuning of $\epsilon$. If $\epsilon$ has to be tuned less severely than $v^2/ M^2$, this can be seen as progress compared to the SM. Let us check whether this is the case. From \eqref{eq:hvevdisc} we can write
\be
\frac{v^2}{M^2} \sim \frac{f^2 \epsilon^2 M^4}{\kappa^2} \ .
\ee
Using \eqref{eq:kappabound} we also have $fg/ M^2 = f \epsilon /M \gg \kappa^2/M^6$. Combing with the above we then find
\be
\frac{v^2}{M^2} \gg \frac{f \epsilon}{M} \quad \Rightarrow \quad \epsilon \ll \frac{1}{f/M} \frac{v^2}{M^2} \ .
\ee
It follows that the tuning required for the model of \cite{Graham:2015cka} is more severe than the equivalent level of tuning required by the SM. (Recall that $f\gtrsim M$.) Thus, when considering the amount of tuning necessary, the setup of \cite{Graham:2015cka} does not correspond to an improvement over the SM.

\subsubsection*{Ad 3) and 4)}
Let us now consider how the amount of tuning of $\epsilon$ has to be adapted when increasing the UV cutoff as $M \rightarrow M' >M$. Initially, we take $f$ and $\kappa$ to be fixed. Then, keeping $v$ constant while increasing the cutoff we have
\be
v \sim \frac{\epsilon M^3 f}{\kappa} = \frac{\epsilon' (M')^3 f}{\kappa} \quad \Rightarrow \quad \epsilon' = \epsilon {\left(\frac{M}{M'}\right)}^3 \ .
\ee
We find that $\epsilon$ has to be modified with the third power of the cutoff while in the SM only a quadratic adjustment is necessary. In this sense the setup of \cite{Graham:2015cka} does not improve on the SM: if we wish to establish a large hierarchy between $v$ and the cutoff scale $M$, a more severe tuning is necessary in the model of \cite{Graham:2015cka} than in the SM. This conclusion would have to be modified if, in addition to $\epsilon$, $\kappa$ and $f$ also need to be adjusted when changing the cutoff. Then an overall milder scaling behavior than quadratic for the three parameters is possible. However, as discussed in section  \ref{sec:smallsteps}, a mild tuning of three parameters can nevertheless be more severe than a stronger tuning of one parameter.

\subsubsection*{Ad 5)}
The model of \cite{Graham:2015cka} is an effective theory with a cutoff. As pointed out above the parameters of the model require re-tuning when this cutoff is raised. Indeed the shift symmetry is broken in a hard way, as evident in the quadratic cutoff dependence of the linear term.

One possibility to include additional particles in a consistent way could be supersymmetry (as originally suggested in~\cite{Graham:2015cka}; for a first attempt in this direction see~\cite{Hardy:2015laa}). One can hope that the role of the cutoff $M$
can then be played by the SUSY breaking scale. The combination of the shift symmetry and SUSY then suggests 
a scheme for adding further fields: they should respect SUSY (up to its breaking scale $M$) and they should respect the shift symmetry up to the required level as well. 
However, a complete and consistent implementation is still an open question.

\subsubsection*{Ad 6) and Ad 7)}
An alternative ansatz towards solving the hierarchy problem would be to devise a mechanism which enforces $v=0$. This is not what the model of \cite{Graham:2015cka} is attempting. Instead of enforcing $v=0$, a solution to the hierarchy problem could start with an extremely small value for $v$. The model of \cite{Graham:2015cka} can indeed produce a vanishingly small value for the Higgs vev by taking $\epsilon$ to be extremely small. But this is by no means easier than producing the correct vev in the first place. In both cases it is the parameter $\epsilon$ that has to be chosen sufficiently small. There is no independent mechanism that could lead to an extremely small Higgs vev.

Furthermore, when decreasing $\epsilon$ for fixed $f$ and $\kappa$ we run into a regime where the Higgs will be trapped in a non-standard vacuum, i.e.~regime (1) of section \ref{dynamics}.

\subsubsection*{Ad 8)}
Compared to the SM the mechanism of \cite{Graham:2015cka} offers one great simplification: the Higgs vev has a simple dependence on the model parameters. 
In particular, the Higgs vev is proportional to the parameter $\epsilon$. This indeed holds in all regularization schemes thereby not restricting the way the physical cutoff is generated.\\
{\bf This is the key advantage of this paradigm over the SM.}

Let us comment in a bit more detail on the nature of the required shift symmetry breaking. Shift symmetries of the type employed in cosmological relaxation are always spontaneously broken with the field $\phi$ as the Goldstone field. If the symmetry was exact the corresponding Goldstone particle would be massless and would have absolutely no potential. Since this is not the case, explicit or anomalous breaking is required. The latter is somewhat more appealing because it lends itself to being a small non-perturbative effect. 
This case of a very weakly violated shift symmetry is parallel to the situation in the Standard Model with respect to scale invariance. We find it therefore hard to argue purely based on the existence of a symmetry. That said we think and hope that this, being a rather different symmetry, will lead to novel and different opportunities for embedding in a more complete theory that explains the smallness of the required symmetry breaking.

\subsubsection*{Ad 9)}
The model of \cite{Graham:2015cka} is self-consistent. Quantum corrections up to the cutoff $M$ will not induce any additional sizeable couplings which are not already present at tree level. However, one can argue how generic the Lagrangian Eq.~(2) of~\cite{Graham:2015cka} really is. Depending on how the axionic shift symmetry is broken, one could also expect operators of the form $g \phi^3$ and $g M \phi^2$. Both terms would modify the dynamics significantly. One hence requires a principle why these terms are vanishingly small or absent.

\section{Conclusions}
In this paper we discuss the mechanism of `Cosmological Relaxation' \cite{Graham:2015cka} in the context of the electroweak hierarchy problem. 

The hierarchy problem is effectively a question of fine-tuning and, consequently, we studied the severity of tuning in `Cosmological Relaxation' models. In the Standard Model a hierarchically small weak scale can be generated by tuning the Higgs mass parameter at the order of $v^2/\Lambda^2$, where $\Lambda$ is the cutoff of the theory and $v=246$ GeV. In relaxation models this tuning is exchanged for a tuning of a different parameter $\epsilon = g / \Lambda$, where $g$ is a scale associated to the complete breaking of an axionic shift symmetry. Our results are as follows: We find that in the original model of `Cosmological Relaxation' \cite{Graham:2015cka} the parameter $\epsilon$ has to be tuned even smaller than $v^2/\Lambda^2$. Furthermore, if we increase the cutoff as $\Lambda \rightarrow \Lambda' > \Lambda$, the tuning in the SM gets worse by a factor of $( \Lambda' / \Lambda)^2$, while it rises by a factor $( \Lambda' / \Lambda)^3$ in the relaxation model. Overall, we find that the required tuning in relaxation models can be much more severe than in the SM.

One attractive feature of models of `Cosmological Relaxation' is, that the Higgs vev $v$ depends on the input parameters in a simple way. In fact, the Higgs vev is a linear function of the small parameter $\epsilon$, i.e.~$v/\Lambda \propto \epsilon$. As $\epsilon$ controls the strength of the breaking of an axionic shift symmetry, demanding a small Higgs vev is now physically equivalent to demanding a weak breaking of the shift symmetry. 
An important novelty of the relaxation model is that this shift symmetry is different from the weakly broken scale invariance of the Standard Model.
A weakly broken shift symmetry is potentially easier to embed into a more complete theory, e.g.~string theory. 
String theory compactifications typically contain axion-like particles in their low-energy spectra \cite{Svrcek:2006yi, Cicoli:2012sz} and also allow for the breaking of axionic shift symmetries by fluxes and branes \cite{Silverstein:2008sg, McAllister:2008hb}. If the axionic shift symmetry can be broken sufficiently weakly in string theory without tuning, the mechanism of `Cosmological Relaxation' would then in turn solve the electroweak hierarchy problem. This is where more work is needed to establish relaxation as a full solution to the hierarchy problem. The possibility of such a weak breaking has been analyzed in the context of axion monodromy inflation in string theory~\cite{Silverstein:2008sg, McAllister:2008hb} and more recently in \cite{Hebecker:2014kva}. In particular the latter study indicates that a very weakly broken shift symmetry requires a tuned cancellation reminiscent of the original problem in the Standard Model.

We also found that for very small values of $\epsilon=g/\Lambda$ electroweak symmetry is broken via the non-perturbative axion potential $\sim |h|\cos(\phi/f)$.
This acts like a source term and therefore a different pattern of electroweak symmetry breaking ensues. For the QCD axion model this is clearly ruled
out. However, for non-QCD models this may lead to an interesting phenomenology. It would be interesting to study if this can be realized in a consistent manner and how this would manifest itself in experiments.

There are many questions which we did not attempt to answer. For `Cosmological Relaxation' to work it has to be embedded in a suitable model of inflation, without introducing the necessity of further tunings. 
Finding such a model which is consistent with all experimental results remains a challenge.

Overall we find that the original model of `Cosmological Relaxation' \cite{Graham:2015cka} contains parameters that need to be tuned to even smaller values than the equivalent in the SM. It remains to be seen whether this is a feature of that particular model or a property of the whole paradigm.
The paramount question then becomes to find ways of benefitting from the improved parameterization so that one can {\emph{explain}} the required small parameters by a suitable mechanism.

\subsubsection*{Acknowledgements}
We thank Martin Bauer, Arthur Hebecker, David E.~Kaplan, Surjeet Rajendran and M.~Spannowsky for very useful discussions.  This
work was supported by the Transregio TR33 ``The Dark Universe''.

\appendix
\section{Appendix}
\label{sec:app}
In this appendix we want to explain the notion that one can always find a parametrization in the Standard Model such that the smallness of the Higgs vev is controlled by a small input parameter.
In the following we closely follow the arguments given in~\cite{Englert:2013gz}.
The general structure of the renormalization group equation for the Higgs mass parameter $\epsilon_{H}=m^{2}_{H}/k^2$ is schematically given by,
\begin{equation}
\label{rgeq}
\partial_{t}\epsilon_{H}=(-2+\eta_{H})\epsilon_{H}+c_{g}g^{2}+c_{\lambda}\lambda.
\end{equation}
$c_{g},c_{\lambda}$ are regularization scheme dependent numbers, and $g$ schematically represents all gauge and fermion contributions.
$\eta_{H}$ is the Higgs anomalous dimension which itself is of the order of $g^2,\lambda^2$ and therefore small in the UV\footnote{Since the U(1) gauge coupling and the Higgs self-coupling are marginally irrelevant this is strictly speaking not quite true in the extreme UV, but this is a different problem and not relevant for our current discussion.}. 

In dimensional regularization $c_{g}=c_{\lambda}=0$. 
For simplicity of our argument let us neglect the running of all couplings $g,\lambda$. We can then easily solve Eq.~\eqref{rgeq},
\begin{equation}
\label{solve}
\epsilon_{H}(k)=\left(\frac{\Lambda}{k}\right)^{2-\eta_{H}}\epsilon_{H}(\Lambda).
\end{equation}
To get the correct order of magnitude for $m^{2}_{H}$ in the IR at $k\sim v$, corresponding to,
\begin{equation}
\epsilon_{H}(k\sim v)=\epsilon_{H}\sim 1,
\end{equation}
one therefore needs to pick
\begin{equation}
\epsilon_{H}(\Lambda)\sim \left(\frac{v}{\Lambda}\right)^{2-\eta_{H}}\ll 1.
\end{equation}
which is exactly the fine-tuning issue in the RG language as discussed above in the discussion of points 3 and 4 around Eq.~\eqref{pick}.
Indeed we can also directly see how a large anomalous dimension $\eta_{H}\approx 2$ can ameliorate the problem.

However, dimensional regularization could be accused of neglecting the all important quadratic divergences. So let us consider the situation $c_{g}\neq 0, c_{\lambda}\neq 0$. 
For constant $g,\lambda$ one can the still quite easily re-write Eq.~\eqref{rgeq} as,
\begin{equation}
\partial_{t}\hat{\epsilon}=(-2+\eta_{H})\hat{\epsilon},
\end{equation}
with
\begin{equation}
\hat{\epsilon}=\epsilon_{H}-\epsilon_{\star},\qquad \epsilon_{\star}=\frac{c_{g} g^2+c_{\lambda}\lambda}{2-\eta_{H}}.
\end{equation}
This obviously has a solution of the same structure as Eq.~\eqref{solve},
\begin{equation}
\label{solve2}
\hat{\epsilon}(k)=\left(\frac{\Lambda}{k}\right)^{2-\eta_{H}}\hat{\epsilon}(\Lambda).
\end{equation}
We obtain,
\begin{equation}
m^{2}_{H}(k\sim v)\sim \hat{\epsilon}(k\sim v) v^2+\epsilon_{\star} v^2\sim \hat{\epsilon}(k\sim v) v^2.
\end{equation}
The latter holds, because $\epsilon_{\star}\ll 1$ for small couplings $g,\lambda$.
Hence, we again require 
\begin{equation}
\label{cond2}
\hat{\epsilon}(k\sim v)\sim 1
\end{equation}
to get the correct electroweak scale.
Thus the fine-tuning is reduced to choosing $\hat{\epsilon}(\Lambda)$ to be exceedingly small. In this sense we can always succeed by choosing a particular parameter to be small\footnote{More generally one can argue for this by defining the theory at a UV fixed point of the renormalization group. The relevant parameters are then the (small) distances from this fixed point in those direction in which the fixed point is UV attractive (hence they get smaller and smaller as we go towards the UV). While the location of the fixed point is not necessarily scheme independent the scaling dimension of the relevant directions are generally thought to be so.}.

One may wonder what happened to the quadratic divergences. Neglecting $\eta_{H}\approx 0$ and inserting Eq.~\eqref{solve2} and using $m^{2}_{H}=\epsilon_{H}(\Lambda)\Lambda^2$ one finds,
\begin{equation}
m^{2}_{H}(k\sim v)\sim  m^{2}_{H}(\Lambda)+\epsilon_{\star}(\Lambda)\Lambda^2\sim m^{2}_{H}(\Lambda)+\frac{c_{g}g^2+c_{\lambda}\lambda}{2}\Lambda^2,
\end{equation}
which contains exactly the expected quadratic terms in $\Lambda$. Choosing $\hat{\epsilon}(\Lambda)\ll 1$ in the UV simply corresponds to choosing an appropriate counterterm for them. 

The all important and (remaining) question is then why one should parametrize the UV theory in terms of $\hat{\epsilon}$ and not in terms of $\hat{\epsilon}_{H}$, the latter not being exceedingly small but requiring to be tuned to a value very close to $\epsilon_{\star}$. In other words
why are we close to the fixed point/phase transition in the first place.

\bibliography{hierarchy}  

\begin{thebibliography}{9}

\bibitem{Graham:2015cka}
  P.~W.~Graham, D.~E.~Kaplan and S.~Rajendran,
  arXiv:1504.07551 [hep-ph].

\bibitem{Abbott:1984qf}
  L.~F.~Abbott,
  Phys.\ Lett.\ B {\bf 150} (1985) 427.

\bibitem{Dvali:2003br}
  G.~Dvali and A.~Vilenkin,
  Phys.\ Rev.\ D {\bf 70} (2004) 063501
  [hep-th/0304043].

\bibitem{Dvali:2004tma}
  G.~Dvali,
  Phys.\ Rev.\ D {\bf 74} (2006) 025018
  [hep-th/0410286].

\bibitem{Espinosa:2015eda}
  J.~R.~Espinosa, C.~Grojean, G.~Panico, A.~Pomarol, O.~Pujolˆs and G.~Servant,
  arXiv:1506.09217 [hep-ph].



\bibitem{Hardy:2015laa}
  E.~Hardy,
  arXiv:1507.07525 [hep-ph].



\bibitem{Patil:2015oxa}
  S.~P.~Patil and P.~Schwaller,
  arXiv:1507.08649 [hep-ph].



\bibitem{Antipin:2015jia}
  O.~Antipin and M.~Redi,
  arXiv:1508.01112 [hep-ph].



\bibitem{Giudice:2008bi}
  G.~F.~Giudice,
  In *Kane, Gordon (ed.), Pierce, Aaron (ed.): Perspectives on LHC physics* 155-178
  [arXiv:0801.2562 [hep-ph]].



\bibitem{Ciafaloni:1996zh}
  P.~Ciafaloni and A.~Strumia,
  Nucl.\ Phys.\ B {\bf 494} (1997) 41
  [hep-ph/9611204].



\bibitem{Dirac:1937ti}
  P.~A.~M.~Dirac,
  Nature {\bf 139} (1937) 323.



\bibitem{Dirac:1938mt}
  P.~A.~M.~Dirac,
  Proc.\ Roy.\ Soc.\ Lond.\ A {\bf 165} (1938) 199.



\bibitem{resonaances}
Cf. post of 8 May 2015 on Resonaances Blog, ``http://resonaances.blogspot.de/''.

\bibitem{Wetterich:1983bi}
  C.~Wetterich,
  Phys.\ Lett.\ B {\bf 140} (1984) 215.



\bibitem{Weinberg:1975gm}
  S.~Weinberg,
  Phys.\ Rev.\ D {\bf 13} (1976) 974.



\bibitem{Susskind:1978ms}
  L.~Susskind,
  Phys.\ Rev.\ D {\bf 20} (1979) 2619.



\bibitem{Dimopoulos:1979es}
  S.~Dimopoulos and L.~Susskind,
  Nucl.\ Phys.\ B {\bf 155} (1979) 237.



\bibitem{Eichten:1979ah}
  E.~Eichten and K.~D.~Lane,
  Phys.\ Lett.\ B {\bf 90} (1980) 125.



\bibitem{tHooft:1979bh}
  G.~'t Hooft,
  NATO Sci.\ Ser.\ B {\bf 59} (1980) 135.

\bibitem{Fayet:1976et}
  P.~Fayet,
  Phys.\ Lett.\ B {\bf 64} (1976) 159.



\bibitem{Fayet:1977yc}
  P.~Fayet,
  Phys.\ Lett.\ B {\bf 69} (1977) 489.



\bibitem{Farrar:1978xj}
  G.~R.~Farrar and P.~Fayet,
  Phys.\ Lett.\ B {\bf 76} (1978) 575.



\bibitem{Fayet:1979sa}
  P.~Fayet,
  Phys.\ Lett.\ B {\bf 84} (1979) 416.



\bibitem{Ramond:1971gb}
  P.~Ramond,
  Phys.\ Rev.\ D {\bf 3} (1971) 2415.



\bibitem{Neveu:1971rx}
  A.~Neveu and J.~H.~Schwarz,
  Nucl.\ Phys.\ B {\bf 31} (1971) 86.



\bibitem{Gervais:1971ji}
  J.~L.~Gervais and B.~Sakita,
  Nucl.\ Phys.\ B {\bf 34} (1971) 632.



\bibitem{Golfand:1971iw}
  Y.~A.~Golfand and E.~P.~Likhtman,
  JETP Lett.\  {\bf 13} (1971) 323
   [Pisma Zh.\ Eksp.\ Teor.\ Fiz.\  {\bf 13} (1971) 452].



\bibitem{Volkov:1973ix}
  D.~V.~Volkov and V.~P.~Akulov,
  Phys.\ Lett.\ B {\bf 46} (1973) 109.



\bibitem{Wess:1974tw}
  J.~Wess and B.~Zumino,
  Nucl.\ Phys.\ B {\bf 70} (1974) 39.



\bibitem{Dimopoulos:1981au}
  S.~Dimopoulos and S.~Raby,
  Nucl.\ Phys.\ B {\bf 192} (1981) 353.



\bibitem{Witten:1981nf}
  E.~Witten,
  Nucl.\ Phys.\ B {\bf 188} (1981) 513.



\bibitem{Dine:1981za}
  M.~Dine, W.~Fischler and M.~Srednicki,
  Nucl.\ Phys.\ B {\bf 189} (1981) 575.



\bibitem{Dimopoulos:1981zb}
  S.~Dimopoulos and H.~Georgi,
  Nucl.\ Phys.\ B {\bf 193} (1981) 150.



\bibitem{Sakai:1981gr}
  N.~Sakai,
  Z.\ Phys.\ C {\bf 11} (1981) 153.



\bibitem{Kaul:1981hi}
  R.~K.~Kaul and P.~Majumdar,
  Nucl.\ Phys.\ B {\bf 199} (1982) 36.



\bibitem{Martin:1997ns}
  S.~P.~Martin,
  Adv.\ Ser.\ Direct.\ High Energy Phys.\  {\bf 21} (2010) 1
   [Adv.\ Ser.\ Direct.\ High Energy Phys.\  {\bf 18} (1998) 1]
  [hep-ph/9709356].



\bibitem{Coleman:1973jx}
  S.~R.~Coleman and E.~J.~Weinberg,
  Phys.\ Rev.\ D {\bf 7} (1973) 1888.



\bibitem{Hempfling:1996ht}
  R.~Hempfling,
  Phys.\ Lett.\ B {\bf 379} (1996) 153
  [hep-ph/9604278].



\bibitem{Meissner:2006zh}
  K.~A.~Meissner and H.~Nicolai,
  Phys.\ Lett.\ B {\bf 648} (2007) 312
  [hep-th/0612165].



\bibitem{Chang:2007ki}
  W.~F.~Chang, J.~N.~Ng and J.~M.~S.~Wu,
  Phys.\ Rev.\ D {\bf 75} (2007) 115016
  [hep-ph/0701254 [HEP-PH]].



\bibitem{Foot:2007as}
  R.~Foot, A.~Kobakhidze and R.~R.~Volkas,
  Phys.\ Lett.\ B {\bf 655} (2007) 156
  [arXiv:0704.1165 [hep-ph]].



\bibitem{Foot:2007iy}
  R.~Foot, A.~Kobakhidze, K.~L.~McDonald and R.~R.~Volkas,
  Phys.\ Rev.\ D {\bf 77} (2008) 035006
  [arXiv:0709.2750 [hep-ph]].



\bibitem{Meissner:2007xv}
  K.~A.~Meissner and H.~Nicolai,
  Phys.\ Lett.\ B {\bf 660} (2008) 260
  [arXiv:0710.2840 [hep-th]].



\bibitem{Iso:2009ss}
  S.~Iso, N.~Okada and Y.~Orikasa,
  Phys.\ Lett.\ B {\bf 676} (2009) 81
  [arXiv:0902.4050 [hep-ph]].



\bibitem{Holthausen:2009uc}
  M.~Holthausen, M.~Lindner and M.~A.~Schmidt,
  Phys.\ Rev.\ D {\bf 82} (2010) 055002
  [arXiv:0911.0710 [hep-ph]].



\bibitem{AlexanderNunneley:2010nw}
  L.~Alexander-Nunneley and A.~Pilaftsis,
  JHEP {\bf 1009} (2010) 021
  [arXiv:1006.5916 [hep-ph]].



\bibitem{Bardeen:1995kv}
  W.~A.~Bardeen,
  FERMILAB-CONF-95-391-T, C95-08-27.3.



\bibitem{Englert:2013gz}
  C.~Englert, J.~Jaeckel, V.~V.~Khoze and M.~Spannowsky,
  JHEP {\bf 1304} (2013) 060
  [arXiv:1301.4224 [hep-ph]].



\bibitem{Randall:1999ee}
  L.~Randall and R.~Sundrum,
  Phys.\ Rev.\ Lett.\  {\bf 83} (1999) 3370
  [hep-ph/9905221].



\bibitem{Randall:1999vf}
  L.~Randall and R.~Sundrum,
  Phys.\ Rev.\ Lett.\  {\bf 83} (1999) 4690
  [hep-th/9906064].



\bibitem{ArkaniHamed:1999dc}
  N.~Arkani-Hamed and M.~Schmaltz,
  Phys.\ Rev.\ D {\bf 61} (2000) 033005
  [hep-ph/9903417].



\bibitem{Abel:2006yk}
  S.~A.~Abel and M.~D.~Goodsell,
  JHEP {\bf 0710} (2007) 034
  [hep-th/0612110].



\bibitem{Cvetic:2009yh}
  M.~Cvetic, J.~Halverson and R.~Richter,
  JHEP {\bf 0912} (2009) 063
  [arXiv:0905.3379 [hep-th]].



\bibitem{Froggatt:1978nt}
  C.~D.~Froggatt and H.~B.~Nielsen,
  Nucl.\ Phys.\ B {\bf 147} (1979) 277.



\bibitem{Seiberg:1993vc}
  N.~Seiberg,
  Phys.\ Lett.\ B {\bf 318} (1993) 469
  [hep-ph/9309335].



\bibitem{Djouadi:1999gv}
  A.~Djouadi, W.~Kilian, M.~Muhlleitner and P.~M.~Zerwas,
  Eur.\ Phys.\ J.\ C {\bf 10} (1999) 27
  [hep-ph/9903229].



\bibitem{Baur:2002rb}
  U.~Baur, T.~Plehn and D.~L.~Rainwater,
  Phys.\ Rev.\ Lett.\  {\bf 89} (2002) 151801
  [hep-ph/0206024].



\bibitem{Baur:2002qd}
  U.~Baur, T.~Plehn and D.~L.~Rainwater,
  Phys.\ Rev.\ D {\bf 67} (2003) 033003
  [hep-ph/0211224].



\bibitem{Baur:2003gpa}
  U.~Baur, T.~Plehn and D.~L.~Rainwater,
  Phys.\ Rev.\ D {\bf 68} (2003) 033001
  [hep-ph/0304015].



\bibitem{Baur:2003gp}
  U.~Baur, T.~Plehn and D.~L.~Rainwater,
  Phys.\ Rev.\ D {\bf 69} (2004) 053004
  [hep-ph/0310056].



\bibitem{Dolan:2012rv}
  M.~J.~Dolan, C.~Englert and M.~Spannowsky,
  JHEP {\bf 1210} (2012) 112
  [arXiv:1206.5001 [hep-ph]].



\bibitem{Baglio:2012np}
  J.~Baglio, A.~Djouadi, R.~Gršber, M.~M.~MŸhlleitner, J.~Quevillon and M.~Spira,
  JHEP {\bf 1304} (2013) 151
  [arXiv:1212.5581 [hep-ph]].



\bibitem{Barr:2014sga}
  A.~J.~Barr, M.~J.~Dolan, C.~Englert, D.~E.~Ferreira de Lima and M.~Spannowsky,
  JHEP {\bf 1502} (2015) 016
  [arXiv:1412.7154 [hep-ph]].



\bibitem{Svrcek:2006yi}
  P.~Svrcek and E.~Witten,
  JHEP {\bf 0606} (2006) 051
  [hep-th/0605206].



\bibitem{Cicoli:2012sz}
  M.~Cicoli, M.~Goodsell and A.~Ringwald,
  JHEP {\bf 1210} (2012) 146
  [arXiv:1206.0819 [hep-th]].



\bibitem{Silverstein:2008sg}
  E.~Silverstein and A.~Westphal,
  Phys.\ Rev.\ D {\bf 78} (2008) 106003
  [arXiv:0803.3085 [hep-th]].



\bibitem{McAllister:2008hb}
  L.~McAllister, E.~Silverstein and A.~Westphal,
  Phys.\ Rev.\ D {\bf 82} (2010) 046003
  [arXiv:0808.0706 [hep-th]].



\bibitem{Hebecker:2014kva}
  A.~Hebecker, P.~Mangat, F.~Rompineve and L.~T.~Witkowski,
  Nucl.\ Phys.\ B {\bf 894} (2015) 456
  [arXiv:1411.2032 [hep-th]].



\end{thebibliography}
\bibliographystyle{JHEP}

\end{document}